\documentclass[
reprint,
superscriptaddress,
preprintnumbers,
nobibnotes,nofootinbib,
amsmath,amssymb,
aps,
floatfix,
]{revtex4-1}

\usepackage{gensymb}
\usepackage{fixmath}
\usepackage{mathrsfs}
\usepackage{graphicx}
\usepackage{dcolumn}
\usepackage{bm}
\usepackage{tabularx}
\usepackage{floatrow}
\usepackage{titlesec}
\usepackage{cancel}
\usepackage{amsmath}
\titleformat*{\section}{\normalsize\bfseries\filcenter}

\titlespacing*{\section}
{0pt}{3.0ex}{1.5ex}

\titlespacing*{\subsection}
{0pt}{2.5ex}{0.5ex}
\titlespacing*{\subsubsection}
{0pt}{2.0ex}{0.5ex}

\usepackage[
  colorlinks=true,
  urlcolor=blue,
  linkcolor=blue,
  citecolor=blue
]{hyperref}
\usepackage{xcolor}
\usepackage{textcomp}
\usepackage{amsmath}
\usepackage{siunitx}
\usepackage[artemisia]{textgreek}
\usepackage[normalem]{ulem}
\usepackage{upgreek}
\usepackage{etoolbox}

\usepackage{graphicx}
\usepackage{graphics}
\usepackage{subfigure}
\usepackage{amsmath}
\usepackage{psfrag}
\usepackage{epsfig}
\usepackage{lineno}
\usepackage{hyperref}
\usepackage{amssymb}
\usepackage{url}
\usepackage{xcolor}
\usepackage[colorinlistoftodos]{todonotes}

\newcolumntype{C}{>{\centering\arraybackslash}X}

\ProvideTextCommandDefault{\textonehalf}{${}^1\!/\!{}_2\ $}
\begin{document}
\title{Proposal for the search for new spin interactions at the micrometer scale using diamond quantum sensors}

\author{P.-H.~Chu$^{\mathsection}$}
\email[Email address:]{pchu@lanl.gov}
\affiliation{Los Alamos National Laboratory, Los Alamos, New Mexico 87545, USA}
\author{N.~Ristoff$^{\mathsection}$}
\affiliation{Center for High Technology Materials, 
University of New Mexico, Albuquerque, NM 87106, USA}
\affiliation{Department of Physics and Astronomy,
University of New Mexico, Albuquerque, NM 87106, USA}
\author{J.~Smits}
\affiliation{Center for High Technology Materials, 
University of New Mexico, Albuquerque, NM 87106, USA}
\affiliation{Department of Physics and Astronomy,
University of New Mexico, Albuquerque, NM 87106, USA}
\author{N.~Jackson}
\affiliation{Center for High Technology Materials, 
University of New Mexico, Albuquerque, NM 87106, USA}
\affiliation{Department of Mechanical Engineering,
University of New Mexico, Albuquerque, NM 87106, USA}
\author{Y.~J.~Kim}
\affiliation{Los Alamos National Laboratory, Los Alamos, New Mexico 87545, USA}
\author{I.~Savukov}
\affiliation{Los Alamos National Laboratory, Los Alamos, New Mexico 87545, USA}
\author{V.~M.~Acosta}
\email[Email address:]{vmacosta@unm.edu}
\affiliation{Center for High Technology Materials, 
University of New Mexico, Albuquerque, NM 87106, USA}
\affiliation{Department of Physics and Astronomy,
University of New Mexico, Albuquerque, NM 87106, USA}
\renewcommand{\thefootnote}{}{\footnote{$\mathsection$ These authors contributed equally to this work.}}
\date{\today}

\begin{abstract}
For decades, searches for exotic spin interactions have used increasingly-precise laboratory measurements to test various theoretical models of particle physics. However, most searches have focused on interaction length scales of $\gtrsim1~{\rm mm}$, corresponding to hypothetical boson masses of $\lesssim0.2~{\rm meV}$. Recently, quantum sensors based on Nitrogen-Vacancy (NV) centers in diamond have emerged as a promising platform to probe spin interactions at the micrometer scale, opening the door to explore new physics at this length scale. Here, we propose experiments to search for several hypothetical interactions between NV electron spins and moving masses. We focus on potential interactions involving the coupling of NV spin ensembles to both spin-polarized and unpolarized masses attached to vibrating mechanical oscillators. For each interaction, we estimate the sensitivity, identify optimal experimental conditions, and analyze potential systematic errors. Using multi-pulse quantum sensing protocols with NV spin ensembles to improve sensitivity, we project new constraints that are $\gtrsim5$ orders-of-magnitude improvement over previous constraints at the micrometer scale. We also identify a spin-polarized test mass, based on hyperpolarized $^{13}$C nuclear spins in a thin diamond membrane, which offers a favorable combination of high spin density and low stray magnetic fields. Our analysis is timely in light of a recent preprint~\cite{rong:2020observation} reporting a surprising non-zero result of µm-scale spin-velocity interactions.
\end{abstract}
\pacs{32..Dk, 11.30.Er, 77.22.-d, 14.80.Va,75.85.+t}
\keywords{axion, dark matter, NV centers, diamond, quantum sensors}

\maketitle

\section{Introduction}
Since the early days of modern physics, it has been established that `spin' is an intrinsic form of angular momentum, carried by elementary particles, that interacts with its environment by the exchange of photons. However, the Standard Model of Particle Physics does not preclude the possibility that spins interact with their environment in other ways. The presence of such ``exotic'' interactions would indicate the exchange of new bosonic particles, an exciting prospect that has been explored for at least $40~{\rm years}$~\cite{Safronova:2018}. 

Among the hypothetical new bosons, the spin-0 `axion' is arguably the most prominent candidate. It has been proposed that the existence of axions could explain the strong Charge-Parity violation~\cite{Peccei:1977} and the source of dark matter~\cite{Duffy:2009}. Axions can couple to fermions through a scalar or pseudoscalar vertex~\cite{Moody:1984}. Similar to the electromagnetic force, these two-fermion potentials can be modeled as Yukawa-type interactions, which depend on the interaction length, the distance between the fermions, and the fermion masses. An analogous two-fermion potential can involve exchange of spin-1 bosons~\cite{Dobrescu:2006}, leading to a dependence of the interaction on the relative velocity between the two fermions. 

The theoretical motivation for probing new spin interactions goes beyond the search for axions. Exotic spin interactions could potentially underlie a number of unexplained particle-physics phenomena including the hierarchy problem, dark energy, and dark photons~\cite{Safronova:2018}. Some hypothetical interactions are not invariant under parity-inversion or time-reversal symmetries. Therefore their observation could provide sources for symmetry violations, which are essential to explain the matter-antimatter asymmetry of the universe~\cite{Canetti:2012}. 

Exotic spin interactions have been experimentally probed over a broad range of interaction-length and boson-mass scales. Numerous constraints have been set from analysis of particle accelerator and astrophysical observations~\cite{Raffelt:2012}. Tabletop experiments have expanded the range of constraints, particularly in the light mass ($\lesssim1~{\rm meV}$) and macroscopic length ($\gtrsim0.2~{\rm mm}$) range. Examples of these experiments include nuclear magnetic resonance spectroscopy~\cite{Garcon:2019, Wu:2019, Lee:2018, su:2021}, spin-polarized torsion pendulum measurements~\cite{Heckel:2008, Terrano:2015, Hunter:2013, Hoedl:2011}, atomic beam spectroscopy~\cite{Ficek:2017, Leefer:2016}, and ultracold molecular spectroscopy~\cite{Altuntas:2018, Roussy:2020}.

Magnetometers based on spin precession can also be used to constrain exotic spin interactions, as the hypothetical interactions affect spin dynamics in a similar manner as magnetic fields. This approach has been pursued for decades using sensitive magnetometers based on, for example, alkali vapor~\cite{Chu:2016, wang:2018, Lee:2018, Kim:2018, Kim:2019sry, chu:2020} and superconducting quantum interference devices~\cite{Ni:1999, Brubaker:2017, gramolin:2021}. 

Recently ``quantum sensors''~\cite{Degen:2017} based on Nitrogen-Vacancy (NV) centers in diamond have emerged as an alternative platform capable of measuring spin interactions down to the sub-micrometer scale. In the last few years, single NV centers have been used to search for several static and velocity-dependent spin interactions. These measurements have set constraints at the micrometer scale, corresponding to constraints on boson masses in the meV range~\cite{Rong:2017,jiao:2020}. 

In late 2020, a preprint was posted that reported the surprising observation of a non-zero coupling between a single NV center and a mechanically-oscillating SiO$_2$ mass~\cite{rong:2020observation}. The analysis suggested the presence of two interaction lengths, $0.4~{\rm \upmu m}$ and $8~{\rm \upmu m}$, implying the existence of two new bosons with masses of $0.5~{\rm eV}$ and $25~{\rm meV}$, respectively. These results have not yet, to our knowledge, been independently replicated, and they potentially contradict constraints from prior observations~\cite{Raffelt:2012}. Nevertheless, the exciting implications call for new measurement schemes to be identified that can replicate, extend, and generalize the findings.

In this Manuscript, we propose experiments to search for several hypothetical interactions between NV electron spin ensembles and micron-scale test masses attached to a vibrating mechanical oscillator. The use of unpolarized test masses allows for replication and generalization of the possible non-zero interaction reported in \cite{rong:2020observation}. However, such spin-velocity interactions can be mediated by either spin-0 or spin-1 bosons. On the other hand, a velocity-dependent spin-spin interaction can only be mediated by exchange of a spin-1 boson~\cite{Fadeev:2019,Leslie:2014}. Thus, we also propose experiments to probe interactions between NV centers and moving spin-polarized test masses, which may provide crucial additional information. In order to minimize systematic errors due to magnetic fields produced by the test mass, we propose to use a spin-polarized test mass based on hyperpolarized $^{13}$C nuclear spins in a thin diamond membrane~\cite{London:2013,Pagliero:2018,Ajoy:2018}. This test mass offers a favorable combination of high spin density ($>1~{\rm nm^{-3}}$) and polarization ($\gtrsim3\%$, controlled by laser light), with a low stray magnetic field due to the small $^{13}$C magnetic moment and membrane geometry. 

\section{Experimental Configuration}
\begin{figure}[htb]
\includegraphics[width=0.7\columnwidth]{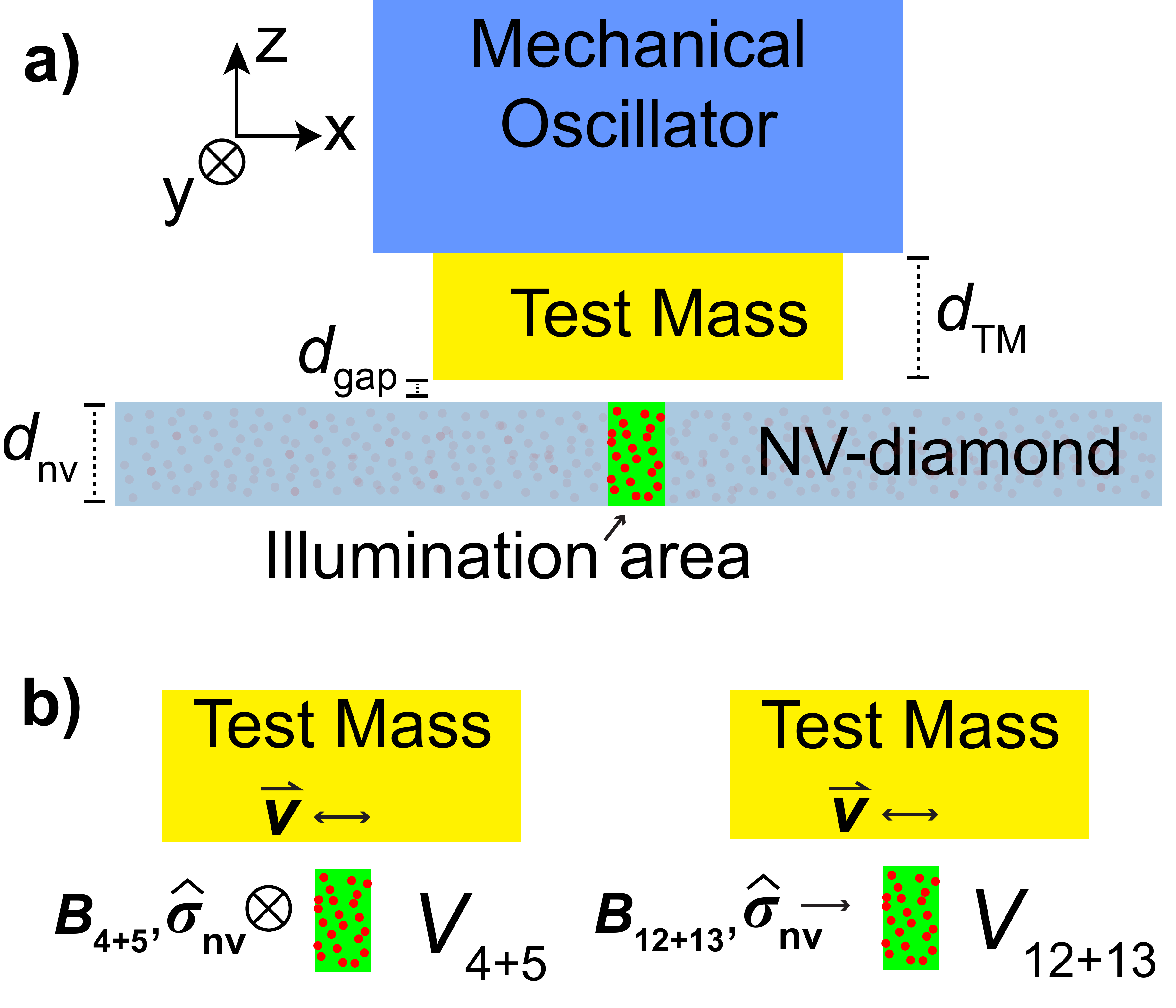}
\caption{\textbf{NV-mechanical oscillator geometry}. (a) Schematic of the experimental geometry. A test mass, attached to a mechanical oscillator, oscillates near an illuminated ensemble of NV centers in a diamond slab. (b) Experimental geometry for each of the two proposed potentials involving unpolarized test masses, $V_{4+5}$ and $V_{12+13}$. Directions of test mass velocity and NV spin polarization are indicated. The approximate direction of the exotic interaction effective magnetic fields is along $\bm{\hat{\sigma}_{\rm nv}}$.}
\label{fig:schematic}
\end{figure}

A schematic of the proposed experimental setup is shown in Fig.~\ref{fig:schematic}. A diamond chip is doped with a near-surface layer of NV centers (layer thickness: $d_{\rm nv}$). A region of the diamond is illuminated by a laser beam (radius: $R_{\rm nv}$) to interrogate a cylindrical-shaped volume of NV centers. A cylindrical test mass (radius: $R_{\rm tm}$, thickness: $d_{\rm tm}$), positioned a distance $d_{\rm gap}$ above the diamond surface, is attached to a mechanical resonator. The resonator is electrically actuated such that the test mass oscillates laterally with a time-dependent displacement $x(t)$ and velocity $v(t)$. Using this setup, we analyze the possibility to probe five hypothetical potentials characterizing two possible spin-velocity interactions ($V_{12+13}$ and $V_{4+5}$, using the notation of~\cite{Dobrescu:2006}) and three velocity-dependent spin-spin interactions ($V_{6+7}$, $V_{14}$, $V_{15}$).

\subsection{Interaction potentials as effective magnetic fields}
\label{sec:v12}
The exotic spin-dependent potentials can be formulated as effective magnetic fields acting on the NV electron spins. As an example, consider one of the simplest possible interactions: a spin-velocity interaction mediated by spin-1 bosons. The interaction can be described by a time-dependent potential, $V_{12+13}(t)$, as:
\begin{align}\label{eq:V12}
    V_{12+13}(t)&=f_{12+13}\frac{\hbar}{8\pi}[\bm{\hat{\sigma}_{\rm nv}}\cdot\bm{v}(t)]\frac{1}{r(t)}e^{-\frac{r(t)}{\lambda}},
\end{align}
where $\bm{\hat{\sigma}_{\rm nv}}$ is the spin unit vector of the NV electron spin, $\bm{v}(t)$ is the test-mass velocity vector, $r(t)$ is the displacement between the NV centers and test mass nucleons, and $\lambda$ is the interaction length. Here $f_{12+13}$ is a dimensionless coupling strength that is related to other physical quantities by $f_{12+13}=4g_A^e g_V^N$, where $g_A^e$ is the axial vector coupling of the NV electron spins and $g_V^N$ is the vector coupling of the nucleons in the test mass~\cite{Leslie:2014}. 

\begin{figure*}[htb]
\includegraphics[width=\textwidth]{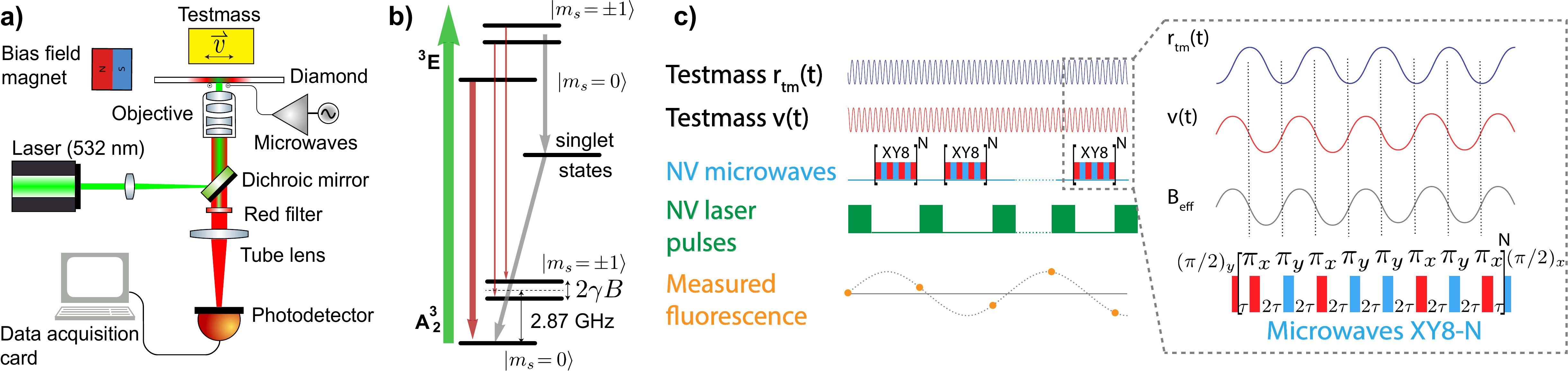}
\caption{\textbf{NV quantum sensors.} (a) Schematic of the NV sensor apparatus. A 532 nm laser beam is used to initialize and read out the NV spin state. The fluorescence is separated from the excitation by a dichroic mirror and a red filter. Microwave pulses are created by an I/Q-modulated microwave generator and delivered to the sensing site by a copper loop. (b) Energy level diagram and optical transitions of the NV center. (c) Pulse sequence used to detect exotic spin interactions. The sequence consists of a series of XY8-N pulse sequences. The time between $\pi$ pulses is equal to half of the oscillation period. The interval between XY8-N measurements determines the aliased frequency of the observed fluorescence signal. }
\label{fig:pulse}
\end{figure*}

Since the potential $V_{12+13}(t)$ involves a dot product with a spin vector, it can be recast as an effective magnetic field interacting with the NV electron spin, with a direction along $\bm{\hat{\sigma}_{\rm nv}}$, given by:
\begin{align}\label{eq:B12} 
B_{12+13}(t) &=f_{12+13}\frac{\rho}{4\pi\gamma}\int_{mass}\bm{\hat{\sigma}_{\rm nv}}\cdot\bm{v}(t)\frac{1}{r(t)}e^{-\frac{r(t)}{\lambda}} d^3r_{\rm tm}\\[3 pt] &\equiv f_{12+13} \frac{1}{4\pi \gamma}\rho F_{12+13}(\lambda,\bm{r}_{\rm tm}(t),\bm{r}_{\rm nv}).
\end{align}
Here $\gamma=28.03~{\rm GHz/T}$ is the NV gyromagnetic ratio, $\rho$ is the test-mass nucleon density, $\bm{r}_{\rm tm}(t)$ is the position vector of the nucleons, $\bm{r}_{\rm nv}$ is the position vector of the NV centers, and $r(t) = |\bm{r}_{\rm tm}(t) - \bm{r}_{\rm nv}|$. To maximize coupling, we assume the test mass oscillates along the $\bm{\hat{\sigma}_{\rm nv}}$ axis, with a velocity:
\begin{align}
    v(t)=2\pi f_m d_1\cos\left( 2\pi f_m t\right),
\end{align}
where $d_1$ is the displacement amplitude of the modulation and $f_m$ is the modulation frequency. Thus, the effective magnetic field, $B_{12+13}(t)$, oscillates with a frequency $f_m$.

Monte Carlo integration methods, similar to those described in Ref.~\cite{Chu:2016,Kim:2019sry}, are used to compute $F_{12+13}(\lambda,\bm{r}_{\rm tm}(t),\bm{r}_{\rm nv})$. Random pairs involving one NV center in the diamond and one nucleon in the test mass are sampled according to their respective geometries and densities. The effective magnetic field is calculated for each NV-nucleon pair and averaged over the NV layer and test mass. This procedure is repeated at different points along the test mass trajectory resulting in a numerical mapping between the field $B_{12+13}(t)$ detected by the NV sensor and the coupling strength, $f_{12+13}$. Effective magnetic fields for the other four potentials we propose to test are calculated in an analogous manner.

\subsection{NV quantum sensor}
We established that the effect of the exotic potentials is to produce an effective oscillating magnetic field at the location of the NV sensor. We now describe how NV sensors can be used to sensitively detect such oscillating magnetic fields. A typical NV sensor apparatus is depicted in Fig.~\ref{fig:pulse}(a). An epifluorescence microscope is used to perform pulsed optically-detected magnetic resonance spectroscopy of NV centers located in a layer near the surface of a diamond chip. The sensor contains billions of NV centers; each serves as a magnetometer by optically detecting changes in their spin transition frequencies, Fig.~\ref{fig:pulse}(b). 

Green laser illumination (typically at a wavelength of 532 nm) drives spin-preserving transitions from the spin-triplet ground state to the spin-triplet excited state. NV centers in the $m_s=0$ excited state preferentially relax to the $m_s=0$ ground state, producing bright fluorescence in the far red (650-750 nm). On the other hand, NV centers in the $m_s=\pm 1$ excited-state levels can either relax down to the $m_s=\pm 1$ ground states, emitting a red photon, or they can non-radiatively relax to the ground state (preferentially to the $m_s=0$ sublevel), via an intersystem crossing involving intermediate singlet states. Due to their higher probability of non-radiative relaxation, NV centers excited from the $m_s=\pm 1$ levels fluoresce less brightly than those excited from the $m_s=0$ level. The same mechanism also produces a high degree of spin polarization in the $m_s=0$ ground-state level under continuous illumination. 

These properties provide an optical method for initializing and detecting the NV spin state. NV spin precession can then be optically detected by applying pulses of microwaves that are resonant with one of the $m_s=0 \leftrightarrow m_s=\pm1$ spin transition frequencies. In our proposed experiments, a bias field (${\sim}10~{\rm mT}$) is applied to lift the degeneracy between the $m_s=\pm 1$ levels and the microwaves are tuned to the $m_s=0 \leftrightarrow m_s=+1$ transition.

The pulse sequence for the proposed measurements is shown in Fig.~\ref{fig:pulse}(c). It is nearly identical to the synchronized readout scheme used for NV-detected nuclear magnetic resonance spectroscopy experiments~\cite{Smits:2019,glenn2018high}. The mechanical oscillator is actuated to continuously vibrate the test mass, producing an oscillating effective magnetic field with a frequency $f_m$. During this time, a series of XY8-N microwave pulse sequences are applied to the NV centers to detect the effective oscillating field. For each XY8-N sequence, a laser pulse is applied to initialize NV centers into $m_s=0$. A microwave $\pi/2$ pulse is applied to generate a superposition state between the $m_s=0$ and $m_s=+1$ levels. Next, a series of microwave $\pi$ pulses is applied to the NV centers with an inter-pulse spacing $2\tau \approx 1/(2f_m)$. In between pulses, the NV spin states freely precess, accumulating phase due to interactions with the environment - in this case due to the effective field created by the exotic potential. Next, a second microwave $\pi/2$ pulse is applied which maps the accumulated phase into a population difference in the $m_s=\{0,1\}$ basis. Finally, a laser pulse is applied and the detected fluorescence rate is proportional to the spin expectation value in the $m_s=\{0,1\}$ basis. 

In analogy to a lock-in amplifier or heterodyne interferometer, the XY8-N sequence acts as a phase-sensitive bandpass filter (center frequency: $f_m$, bandwidth: ${\sim}f_m/[4N]$), suppressing noise outside the filter’s pass-band. The fluorescence rate observed at the end of each sequence is proportional to the instantaneous value of the effective magnetic field at the beginning of the sequence. By performing a series of sequential XY8-N measurements, an aliased version of the effective field trace is obtained. Fourier analysis of the time trace can be used to observe the frequency and phases of signals as well as characterize the noise.

\subsection{Sensitivity}
The minimum detectable field, $\delta B_{\rm min}$, for the synchronized readout scheme is:
\begin{align}
\label{eq:dbmin}
    \delta B_{\rm min} = \frac{1}{4 \gamma_{nv} C}\frac{1}{\sqrt{n V_{\rm sens}\,\eta\,\delta\, t\, \tau_{tot}}},
\end{align}
where $\gamma_{nv}$=28.03 GHz/T is the NV gyromagnetic ratio, $C$ is the contrast (relative difference in fluorescence rate between $m_s=0$ and $m_s=1$), $n$ is the NV density, $V_{\rm sens}$ is the NV sensor volume, $\eta$ is the probability to detect a photoelectron per NV center per readout, $\delta$ is the free precession duty cycle (the fraction of the total measurement time that NV phase is freely accumulating), $\tau_{\rm tot}$ is the NV phase accumulation time during a single XY8-N sequence, and $t$ is the total measurement time. Typical experimental values can reach $C \approx 0.03$, $n \approx 10^6~{\rm \upmu m^{-3}}$, $\eta \approx 0.05$, $\delta \approx 0.8$, and $\tau_{\rm tot} \approx 17~{\rm \upmu s}$~\cite{Smits:2019,glenn2018high}. With these values, we can compute a time- and volume-normalized sensitivity, $\delta B_{\rm min}\sqrt{V_{\rm sens} t} \approx 3.7 \times 10^{-10}~{\rm T\, s^{1/2}\, \upmu m^{3/2}}$, which we use in the calculations in this paper. This sensitivity is an optimistic target, but it is not far from what has been already realized in experiments. For example, in \cite{glenn2018high}, the authors report a time- and volume-normalized sensitivity sensitivity of $\delta B_{\rm min}\sqrt{V_{\rm sens} t} \approx 1.6 \times 10^{-9}~{\rm T\, s^{1/2}\, \upmu m^{3/2}}$ for AC magnetic fields oscillating at 3 MHz. This is within a factor of $5$ of our target sensitivity and improvements are expected by reducing microwave phase noise and other pulse errors.

In the following sections, we will discuss how this experimental setup can be used to constrain spin-velocity and velocity-dependent spin-spin interactions. 

\section{Interactions using unpolarized test masses}
\label{sec:interactions}
\subsection{Experimental design optimization}
\label{sec:design}
We first consider interactions with test masses that have no net spin polarization. We consider geometries optimized for constraining interactions at three length scales, $\lambda=50,~5,~{\rm and}~0.5~{\rm \upmu m}$. The geometry is shown in Fig.~\ref{fig:schematic} and the experimental parameters are shown in Tab.~\ref{tab:v12}. To optimize the experimental parameters for each value of $\lambda$, we define a figure of merit, $S_i= B_i/\delta B_{\rm{min}}$, where $B_i$ is the effective magnetic field from potential $V_i$ and $\delta B_{\rm{min}}$ is given by Eq.~\eqref{eq:dbmin}. 

\begin{figure}[htb]
    \centering
    \includegraphics[width=\columnwidth]{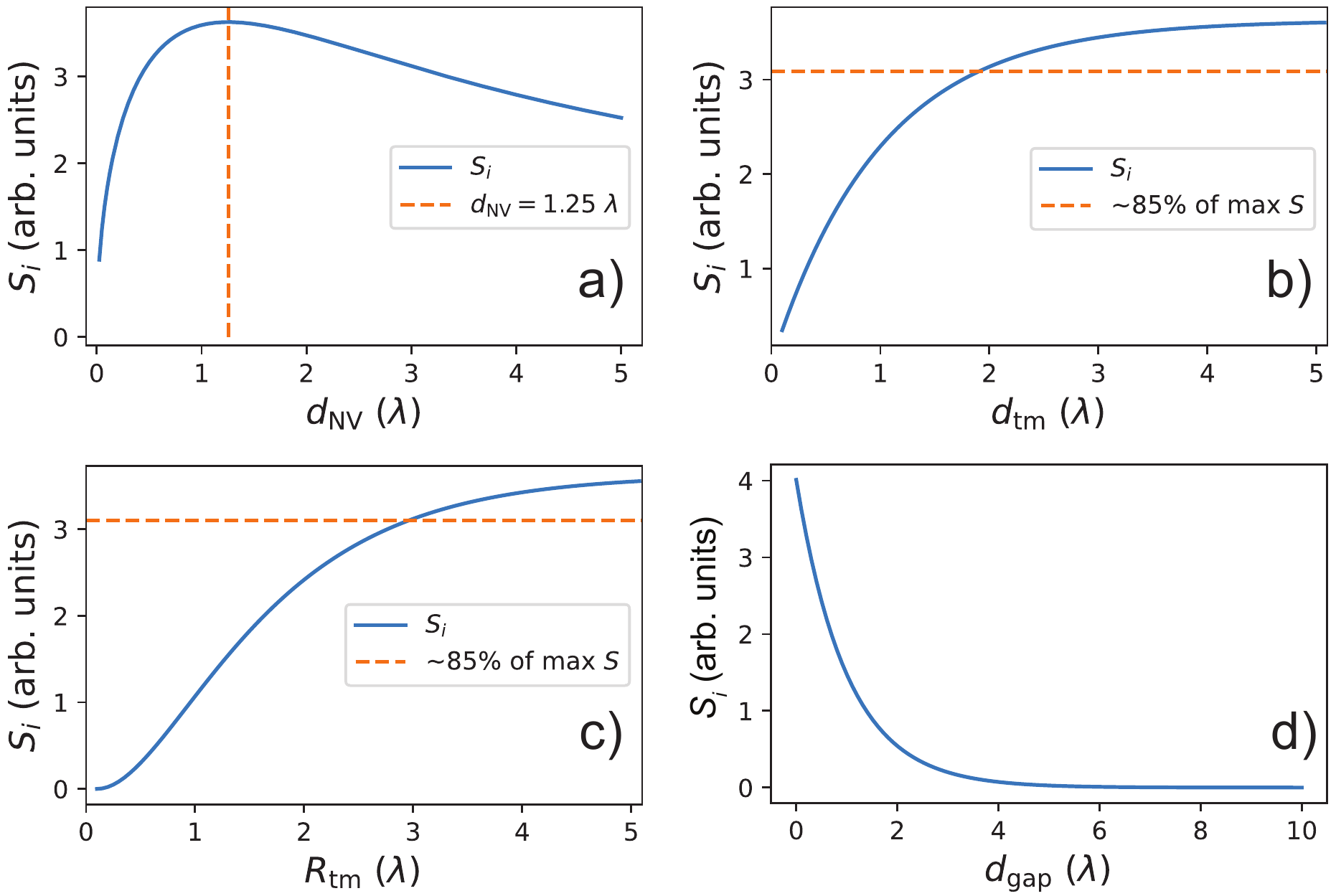}
    \caption{\textbf{Optimization curves.} Sensitivity figure of merit, $S_i$ as a function of a) NV layer thickness, $d_{\rm nv}$, b) test mass thickness, $d_{\rm tm}$, c) test mass radius, $R_{\rm tm}$, and d) NV-test mass standoff, $d_{\rm gap}$. }
\label{fig:geometryopt}
\end{figure}

\begin{table}[htb]
\caption{\textbf{Parameters used to probe interactions with unpolarized test mass, $\bm{V_{12+13}}$ and $\bm{V_{4+5}}$.} $\rho$ is the nucleon density of the test mass. We assume quartz for our calculations, but other materials are discussed in Sec.~\ref{subsec:testmass}. $R_{\rm tm}$ is the radius of the test mass. $d_{\rm tm}$ is the thickness of the test mass. $d_{\rm nv}$ is the thickness of the diamond. $A_{\rm nv}$ is the interrogation area of NV centers in the diamond. $d_{\rm gap}$ is the minimum gap between the test mass and diamond sensor. $d_1$ is the peak displacement, $f_m$ is the modulation frequency, and $v$ is the peak velocity of the test mass oscillation. $\delta B_{\rm min}$ is the estimated magnetic sensitivity for $t=1~{\rm s}$.}
\begin{center}
\begin{tabular}{ c | c | c | c}
Target $\lambda$ & $50~{\rm \upmu m}$ & $5~{\rm \upmu m}$ & $0.5~{\rm \upmu m}$\\ 
\hline 
$\rho$ &  $1.6\times 10^{30}~{\rm m^{-3}}$ & $1.6\times 10^{30}~{\rm m^{-3}}$ & $1.6\times 10^{30}~{\rm m^{-3}}$\\ 
\hline
$R_{\rm tm}$ & $150~{\rm \upmu m}$ & $150~{\rm \upmu m}$ & $150~{\rm \upmu m}$ \\ 
\hline
$d_{\rm tm}$ & $100~{\rm \upmu m}$ & $100~{\rm \upmu m}$ & $100~{\rm \upmu m}$\\ 
\hline
$d_{\rm nv}$ & $62.5~{\rm \upmu m}$ & $6.25~{\rm \upmu m}$ & $0.625~{\rm \upmu m}$\\ 
\hline
$A_{\rm nv}$ & $\pi \times (25~{\rm \upmu m})^2$ & $\pi \times (25~{\rm \upmu m})^2$ & $\pi \times (25~{\rm \upmu m})^2$\\ 
\hline
$d_{\rm gap}$ & $5~{\rm \upmu m}$ & $0.5~{\rm \upmu m}$ & $0.2~{\rm \upmu m}$\\ 
\hline
$d_1$ & $0.75~{\rm \upmu m}$  & $0.75~{\rm \upmu m}$ & $0.75~{\rm \upmu m}$\\ 
\hline
$f_m$ & $1~{\rm MHz}$ & $1~{\rm MHz}$ & $1~{\rm MHz}$ \\
\hline
$v$ & $4.7~{\rm m/s}$ & $4.7~{\rm m/s}$ & $4.7~{\rm m/s}$ \\
\hline
$\delta B_{\rm min}$ & $1~{\rm pT}$ & $3~{\rm pT}$ & $10~{\rm pT}$\\
\hline
\end{tabular}
\end{center}
\label{tab:v12}
\end{table}

For all interaction potentials, the figure of merit is maximized when the test mass has the highest possible peak velocity. Here, we optimistically select the test mass oscillation to have an amplitude of $d_1=0.75~{\rm \upmu m}$ and a frequency of $1~{\rm MHz}$, resulting in a peak velocity $v=4.7~{\rm m/s}$. Design considerations for realizing such a mechanical oscillator are described in Sec.~\ref{sec:mechanics}. All interaction effective fields scale linearly with $v$, so if the peak velocity is lower, the constraints estimated here can be scaled accordingly.

To select the remaining experimental parameters, we start by optimizing the NV layer thickness, $d_{\rm nv}$, while holding the test mass dimensions ($d_{\rm tm}$, $R_{\rm tm}$) and standoff ($d_{\rm gap}$) fixed. We take the limit that the NV illumination cross section $A_{\rm nv}\rightarrow0$, so that the integration over $d_{\rm nv}$ occurs in one dimension. We find that for all potentials considered here, the optimal NV layer thickness is $d_{\rm nv}\approx1.25~\lambda$, Fig.~\ref{fig:geometryopt}(a). 

Next, we optimize the test mass dimensions ($d_{\rm tm},R_{\rm tm}$). The maximum figure of merit occurs when $d_{\rm tm},R_{\rm tm}\rightarrow\infty$. However, for $d_{\rm tm}=2\lambda$ and $R_{\rm tm}=3\lambda$, the figure of merit is already a factor of ${\sim}0.85$ of its maximal value, Figs.~\ref{fig:geometryopt}(a,b). Since $\lambda=50~\rm{\upmu m}$ is the largest interaction length we consider, we set $d_{\rm tm}=100~{\rm \upmu m}$ and $R_{\rm tm}=150~{\rm \upmu m}$ for all cases, for convenience. 

Due to the exponential decay terms in each of the hypothetical interactions, the maximum figure of merit occurs when the NV-test mass gap approaches $d_{\rm gap}\rightarrow0$. Figure \ref{fig:geometryopt}(d) shows the dependence of the figure of merit on $d_{\rm gap}$. To simplify the above optimization procedure for $d_{\rm tm}$, $R_{\rm tm}$, and $d_{\rm nv}$, we set $d_{\rm gap} = 0.1\lambda$. However, for the target $\lambda=0.5~\rm{\mu m}$, we set $d_{\rm gap}=0.2~\rm{\mu m}$ for experimental practicality. These choices of $d_{\rm gap}$ only reduce the figure of merit by a factor of $\lesssim1.4$ from the maximal value.

Finally, we optimize the NV sensor illumination area, $A_{\rm nv}$. Here, there is a competition between optimizing the sensitivity $\delta B_{\rm min}$ ($A_{\rm nv}\rightarrow\infty$) and maximizing the interaction effective field $B_i$ ($A_{\rm nv}\rightarrow0$). Considering that the test mass radius has already been set to $R_{\rm{tm}}=150~{\rm \upmu m}$, we find that setting the NV illumination area to be $A_{\rm nv}=\pi\times(25~{\rm \upmu m})^2$ is a reasonable compromise and results in a reduction of the interaction signal by only a factor of $\lesssim1.1$ compared to the $A_{\rm nv}=0$ case.

\subsection{Interaction \texorpdfstring{$\bm{V_{12+13}}$}{}}
\label{sec:12}
With the experimental parameters optimized, we now return to consider the interaction $V_{12+13}$ introduced in Sec.~\ref{sec:v12}. The potential is given in Eq.~\eqref{eq:V12}, the effective magnetic field $B_{12+13}$ is given in Eq.~\eqref{eq:B12}, and the coupling strength $f_{12+13}$ is defined in Sec.~\ref{sec:v12}. Using the geometry in Fig.~\ref{fig:schematic} and the optimized values in Tab.~\ref{tab:v12}, we calculated the minimum detectable coupling constant $f_{12+13}$ as a function of $\lambda$. The curves are obtained by setting the effective field $B_{12+13}$ equal to the NV minimum detectable field $\delta B_{\rm min}$ after $t=10^4~{\rm s}$ of averaging. Note that this and all other exclusion estimates presented in this work represent an ideal lower bound. They neglect potential sources of systematic error, which are discussed separately in Sec.~\ref{sec:systematics}.

\begin{figure}[htb]
\includegraphics[width=\columnwidth]{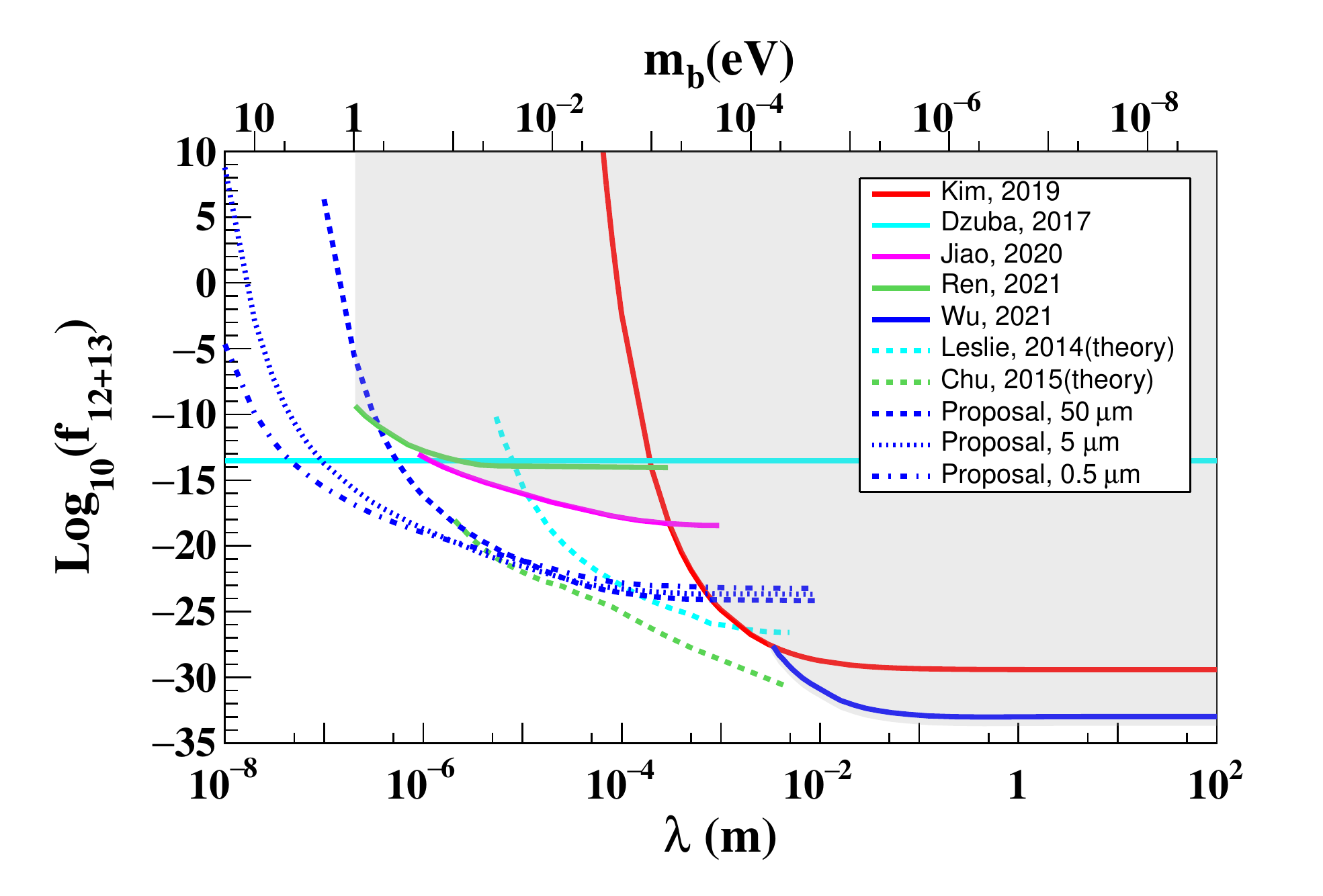}
\caption{\textbf{Exclusion plot for interaction $\bm{V_{12+13}}$.} Exclusion region of electron-nucleon coupling $f_{12+13}$ by previous experiments: Kim, 2019~\cite{Kim:2019sry}, Dzuba, 2017~\cite{Dzuba:2017puc}, Jiao, 2020~\cite{jiao:2020}, Ren, 2021~\cite{Ren:2021} and Wu, 2021~\cite{Wu:2021flk}. The minimum detectable $f_{12+13}$ for our proposed experimental geometries is shown in dashed blue. Theoretical proposals for two other technologies are also shown: Chu, 2015~\cite{Chu:2015tha} and Leslie, 2014~\cite{Leslie:2014}.}
\label{fig:sensitivity_12}
\end{figure}

Figure~\ref{fig:sensitivity_12} shows the minimum detectable coupling constant for each of the 3 different design optimizations in Tab.~\ref{tab:v12}.  As can be seen, while we set a target value of $\lambda$ for each geometry, each experiment is still sensitive to a broad range of interaction lengths. 

Present constraints of $V_{12+13}$ with electrons are set by atomic magnetometers~\cite{Kim:2019sry}, the torsion pendulum~\cite{Heckel:2008hw}, the atomic parity non-conservation experiment~\cite{Dzuba:2017puc}, magnetic force microscopy~\cite{Ren:2021}, and single NV centers~\cite{jiao:2020}. Our proposed experiment could improve on existing experimental constraints by $\gtrsim5$ orders of magnitude in the $1\mbox{-}300~{\rm \upmu m}$ interaction length range. Other proposals for experiments based on paramagnetic insulators~\cite{Chu:2015tha} and mechanical torsion oscillators~\cite{Leslie:2014} promise better performance for $\lambda \gtrsim 100~{\rm \upmu m}$, but they cannot access the smaller interaction lengths that our proposed measurements would reach.

\subsection{Interaction \texorpdfstring{$\bm{V_{4+5}}$}{}}
\label{sec:v4}
The second spin-velocity interaction we analyze is the one studied in Ref.~\cite{rong:2020observation}, which claimed a surprising non-zero coupling. The $V_{4+5}$ potential is written as:
\begin{align}
    V_{4+5} = -f_{4+5}\frac{\hbar^2}{8\pi m_e c}[\bm{\hat{\sigma}_{\rm nv}}\cdot (\bm{v}\times \bm{\hat{r}})](\frac{1}{\lambda r}+\frac{1}{r^2})e^{-\frac{r}{\lambda}},
\end{align}
and the resulting effective magnetic field is given by:
\begin{align}
B_{4+5} &= -f_{4+5}\frac{\hbar\rho}{4\pi m_e c}\int_\text{mass} [\bm{\hat{\sigma}_{\rm nv}}\cdot (\bm{v}\times \bm{\hat{r}})]\notag\\
&\times\left(\frac{1}{\lambda r}+\frac{1}{r^2} \right) e^{-r/\lambda} d^3r_{\rm tm},
\end{align}
where $\bm{\hat{r}} = \bm{r}/r$ is the unit vector in the direction between an NV center electron and a test mass nucleon, and $m_e$ is the mass of an electron. Here and for the remainder of the manuscript, we omit the time-dependence in expressions for brevity. The coupling strength  $f_{4+5}$ is defined as the product of the scalar coupling ($S$) of the NV electron spin and the scalar coupling of the test mass nucleon, or the vector coupling ($V$) of the NV electron spin and the vector coupling of the test mass nucleon. In symbolic form, $f_{4+5}=g_S^e g_S^N$ or $f_{4+5}=g_V^e g_V^N$ for scalar and vector coupling, respectively~\cite{Leslie:2014}.

\begin{figure}[htb]
\includegraphics[width=\columnwidth]{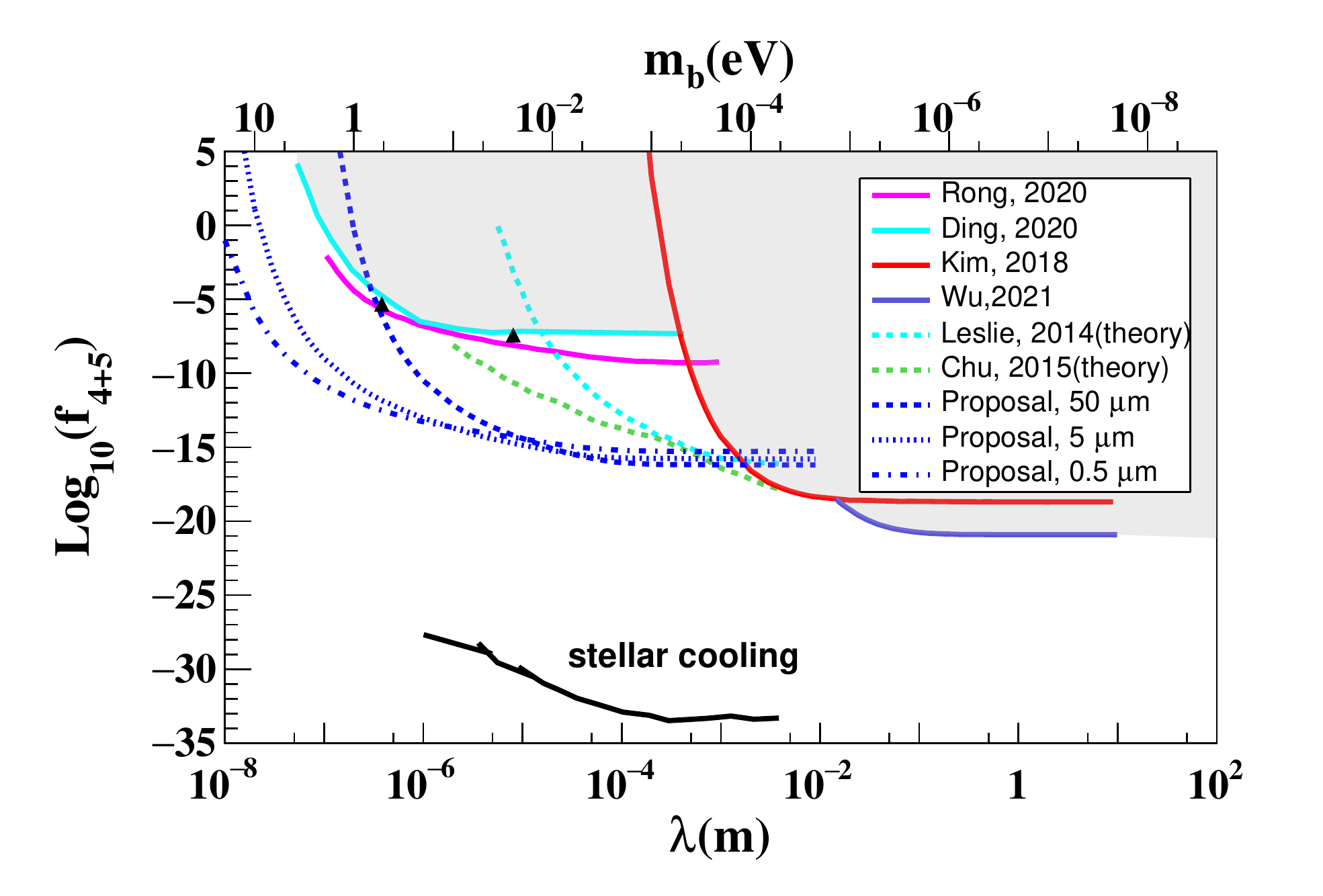}
\caption{\textbf{Exclusion plot for interaction $\bm{V_{4+5}}$.} Exclusion region of electron-nucleon coupling $f_{4+5}$ by experiments Kim, 2018~\cite{Kim:2018}, Ding, 2020~\cite{Ding:2020mic} and Wu, 2021~\cite{Wu:2021flk}. The non-zero result from Rong, 2020~\cite{rong:2020observation}, is shown as black dots, and the sensitivity of that experiment is shown as a solid magenta line. The result from “stellar cooling” observations is shown in solid black. It excludes the result in Ref.~\cite{rong:2020observation} by >25 orders of magnitude, but may be considered a less direct measurement. The minimum detectable $f_{4+5}$ for each of our proposed experimental geometries is shown in dashed blue. Theoretical proposals for two other technologies are also shown: Chu, 2015~\cite{Chu:2015tha} and Leslie, 2014~\cite{Leslie:2014}.}
\label{fig:sensitivity_4}
\end{figure}
The design for the $V_{4+5}$ experimental configuration is shown in Fig.~\ref{fig:schematic}. The parameters (Tab.~\ref{tab:v12}) are the same as $V_{12+13}$ but the test mass motion is perpendicular to the NV spin orientation.

Figure~\ref{fig:sensitivity_4} shows the minimum detectable coupling strength of $f_{4+5}$ for the proposed experiments alongside previous experiments. The black triangular points are the positions of new hypothetical bosons estimated by \cite{rong:2020observation}, and the magenta curve is the reported sensitivity of that experiment. The previous experimental constraints of $V_{4+5}$ are shown as the gray area, set by atomic magnetometers~\cite{Kim:2018, Wu:2021flk}, cantilevers~\cite{Ding:2020mic} and the torsion pendulum ~\cite{Heckel:2008hw}. The stellar cooling limit~\cite{Leslie:2014} is derived with the experimental limit of $g_S^N$ and the astrophysical axion bounds $g_S^e < 1.3 \times 10^{-14}$~\cite{Raffelt:2012}. However, the stellar cooling curve is considered to be an “indirect” measurement of this coupling. The process to generate the curve involves combining results from different experiments, and the uncertainty propagation is not straightforward \cite{Raffelt:2012}. Therefore, a direct measurement of this interaction is still valuable. We should also notice that the stellar cooling limit is far below the claimed discovery which might suggest new systematics for $V_{4+5}$ with the diamond magnetometer. The sensitivity of our design is more than 5 orders more than in Ref.~\cite{rong:2020observation}, which can be used to examine the discovery. Other possible approaches for probing $V_{4+5}$ include the torsional oscillator~\cite{Leslie:2014} and the paramagnetic insulators~\cite{Chu:2015tha}, covering somewhat longer length scales.

\section{Interactions with spin-polarized test masses}
The next set of hypothetical spin interactions we consider are those involving a velocity-dependent spin-spin interaction. Setting new constraints on these interactions requires a test mass that has a high density of polarized spins. The magnetic moments of test mass spins are absent from expressions for the hypothetical potentials, but they produce a systematic effect on the magnetic field detected by the NV electron spins. Thus, we aim to minimize the magnetic fields originating from the magnetized test mass when attempting to detect these interactions.

\subsection{Experimental design}
\begin{figure}[htb]
    \centering
    \includegraphics[width=.8\columnwidth]{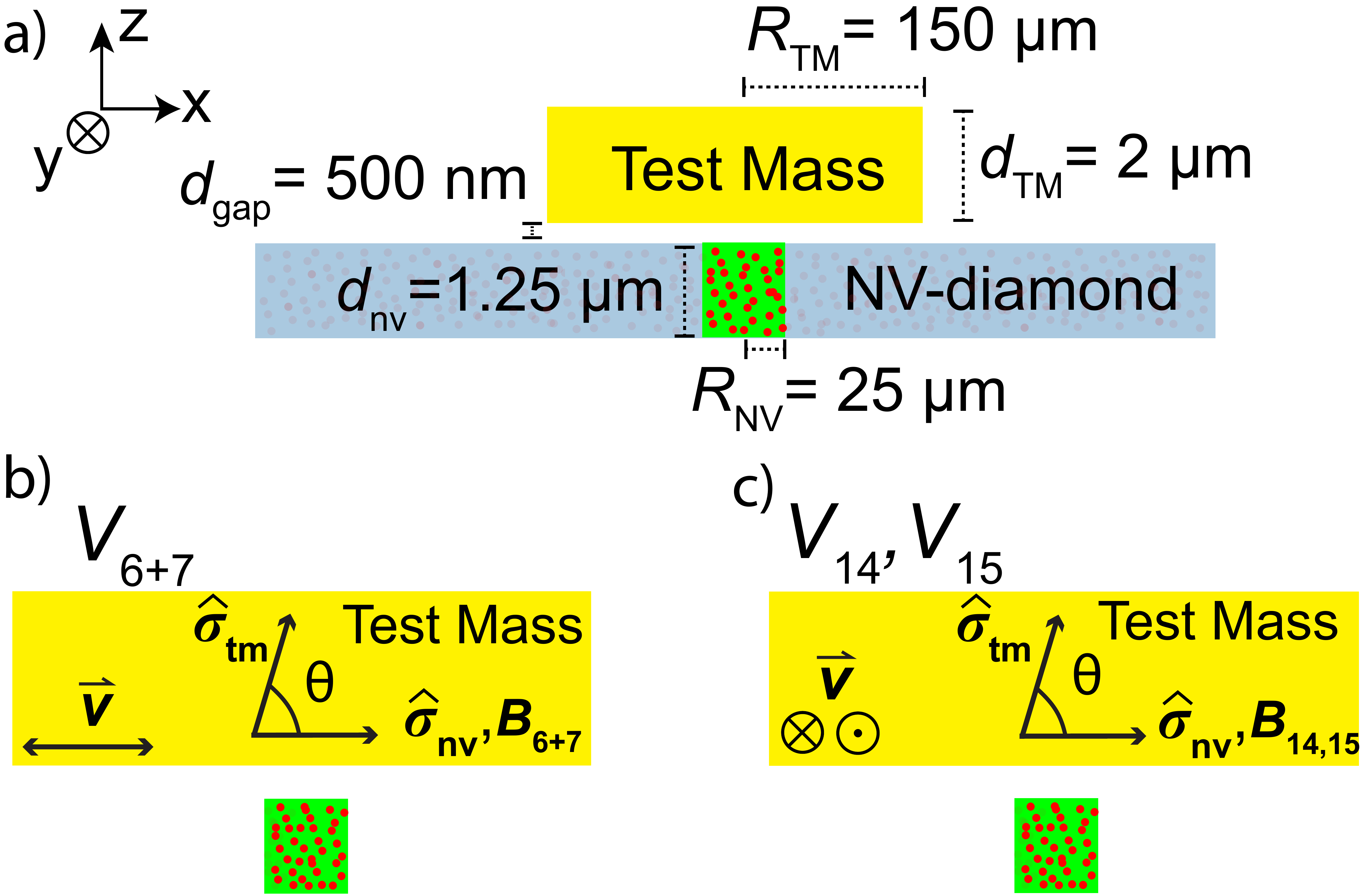}
    \caption{\textbf{Geometry used to probe velocity-dependent spin-spin interactions, $\bm{V_{6+7}}$, $\bm{V_{14}}$, and $\bm{V_{15}}$}. a) Geometry and relevant dimensions for proposed experiments with spin-polarized test masses. b) Experimental geometry for potential $V_{6+7}$. c) Experimental geometry for potentials $V_{14}$ and $V_{15}$. For b) and c), the approximate direction of the exotic interaction effective magnetic fields is along $\bm{\hat{\sigma}_{\rm nv}}$ for $\theta=80\degree$.}
\label{fig:angle_adjust}
\end{figure}

The experimental parameters used to probe $V_{6+7}$, $V_{14}$ and $V_{15}$ are shown in Tab.~\ref{tab:v6+7} and illustrated in Fig.~\ref{fig:angle_adjust}. The experimental parameters are optimized in a similar way as described in Sec.~\ref{sec:design}. Here, we optimize for a single interaction length, $\lambda=1~{\rm \upmu m}$. We first perform the same optimization procedure as before to find $d_{\rm nv}=1.25\lambda$, $R_{\rm tm}=150~{\rm \upmu m}$, and $A_{\rm nv}=\pi\times(25~{\rm \upmu m})^2$. The mechanical oscillator trajectory is also assumed to have the same amplitude, frequency, and peak velocity as in Sec.~\ref{sec:interactions}.  In contrast, however, $d_{\rm tm}$ is decreased to $2~{\rm \upmu m}$ to reduce the undesired stray magnetic signals from the magnetized test mass. In addition, $d_{\rm gap}$ is limited to $0.5~{\rm \upmu m}$ so that undesired stray magnetic signals remain as uniform as possible as the test mass moves.

\begin{table}[h]
\caption{\textbf{Parameters used to probe velocity-dependent spin-spin interactions: $\bm{V_{6+7}}$, $\bm{V_{14}}$ and $\bm{V_{15}}$}. $\rho$ is the density of excess polarized nuclear spins within the test mass (Sec.~\ref{sec:pulsepol}). $R_{\rm tm}$ is the radius of the test mass. $d_{\rm tm}$ is the thickness of the test mass. $d_{\rm nv}$ is the thickness of the diamond. $A_{\rm nv}$ is the interrogation area of NV centers in the diamond.  $d_{\rm gap}$ is the minimum gap between the test mass and diamond sensor. $d_1$ is the peak displacement, $f_m$ is the modulation frequency, and $v$ is the peak velocity of the test mass oscillation. $\delta B_{\rm min}$ is the estimated magnetic sensitivity for $t=1~{\rm s}$.}
\begin{center}
\begin{tabular}{ c | c}
Target $\lambda$ & $1~{\rm \upmu m}$\\ 
\hline 
$\rho$ &  $5\times 10^{25}~{\rm m^{-3}}$ \\ 
\hline
$R_{\rm tm}$ & $150~{\rm \upmu m}$ \\ 
\hline
$d_{\rm tm}$ & $2~{\rm \upmu m}$ \\  
\hline
$d_{\rm nv}$ & $1.25~{\rm \upmu m}$ \\ 
\hline
$A_{\rm nv}$ & $\pi \times (25~{\rm \upmu m})^2$ \\ 
\hline
$d_{\rm gap}$ & $0.5~{\rm \upmu m}$ \\ 
\hline
$d_1$ & $0.75~{\rm \upmu m}$ \\ 
\hline
$f_m$ & $1~{\rm MHz}$ \\
\hline
$v$ & $4.7~{\rm m/s}$ \\
\hline
$\delta B_{\rm min}$ & $6~{\rm pT}$ \\
\hline
\end{tabular}
\end{center}
\label{tab:v6+7}
\end{table}

In order to maximize sensitivity to the hypothetical interactions, while suppressing undesired magnetic signals from the magnetized test mass, we propose to use a test mass containing optically-hyperpolarized $^{13}$C nuclear spins in a thin diamond membrane. The spin-polarized test mass is described in Sec.~\ref{sec:pulsepol}. For sensitivity calculations, we assume an excess polarized-nuclear-spin density of $\rho=5\times 10^{25}~{\rm m^{-3}}$.

\subsection{Interaction \texorpdfstring{$\bm{V_{6+7}}$}{}}
\label{sec:v6}
The interaction $V_{6+7}$ can be mediated by spin-1 bosons~\cite{Dobrescu:2006,Leslie:2014}. The potential is defined as:
\begin{eqnarray}
\label{eq:pot_v67}
V_{6+7} &=  -f_{6+7}\frac{\hbar^2}{8\pi m_{e}c}\left[(\bm{\hat{\sigma}_{\rm nv}}\cdot\bm{v})(\bm{\hat{\sigma}_{\rm tm}}\cdot\bm{\hat{r}})\right]\notag\\
&\times\left(\frac{1}{\lambda r}+\frac{1}{r^{2}}\right)e^{-r/\lambda},
\end{eqnarray}

where $\bm{\hat{\sigma}_{\rm nv}}$ is the NV center electron spin unit vector and $\bm{\hat{\sigma}_{\rm tm}}$ is the spin unit vector of the polarized test mass. 

The coupling strength is the product of the axial vector coupling of the NV center electron and the vector coupling of the polarized mass~\cite{Leslie:2014,Hunter:2014} or vice versa. The coupling strength $f_{6+7}$ is defined as $f_{6+7} \equiv {g_V^{e} g_A^{N}}/{2} + {g_A^{e} g_V^{N}}({m_{e}}/{m_{N}})/{2}$, where $g_A^{e}$ is NV electron axial vector coupling, $g_A^{N}$ is the test mass nucleon axial vector coupling, $g_V^{e}$ is the NV electron vector coupling, $g_V^{N}$ is the test mass vector coupling, $m_e$ is the electron mass, and $m_N$ is the nucleon mass.

The effective magnetic field due to $V_{6+7}$ is: 
\begin{align}
B_{6+7} & =  -f_{6+7}\frac{\hbar\rho}{4\pi\gamma m_{e}c}\notag\\
&\times\int_{mass}\left[(\bm{\hat{\sigma}_{\rm nv}}\cdot\bm{v})(\bm{\hat{\sigma}_{\rm tm}}\cdot\bm{\hat{r}})\right]\left(\frac{1}{\lambda r}+\frac{1}{r^{2}}\right)e^{-r/\lambda} d^3r_{\rm tm}.
\label{eq:beff_v67}
\end{align}
The experimental geometry is shown in Fig.~\ref{fig:angle_adjust}(b). The nuclear spin direction of the test mass, $\bm{\hat{\sigma}_{\rm tm}}$, is oriented at an angle $\theta\approx80\degree$ with respect to the NV electron spin direction, $\bm{\hat{\sigma}_{\rm nv}}$. This represents a compromise between maximizing the NV sensor sensitivity, while minimizing the effect of stray magnetic fields from the test mass, as discussed in Secs.~\ref{sec:pulsepol},\ref{sec:gradients}. The movement of the polarized test mass is parallel to $\bm{\hat{\sigma}_{\rm nv}}$. 

\begin{figure}[hbt]
\includegraphics[width=1\textwidth]{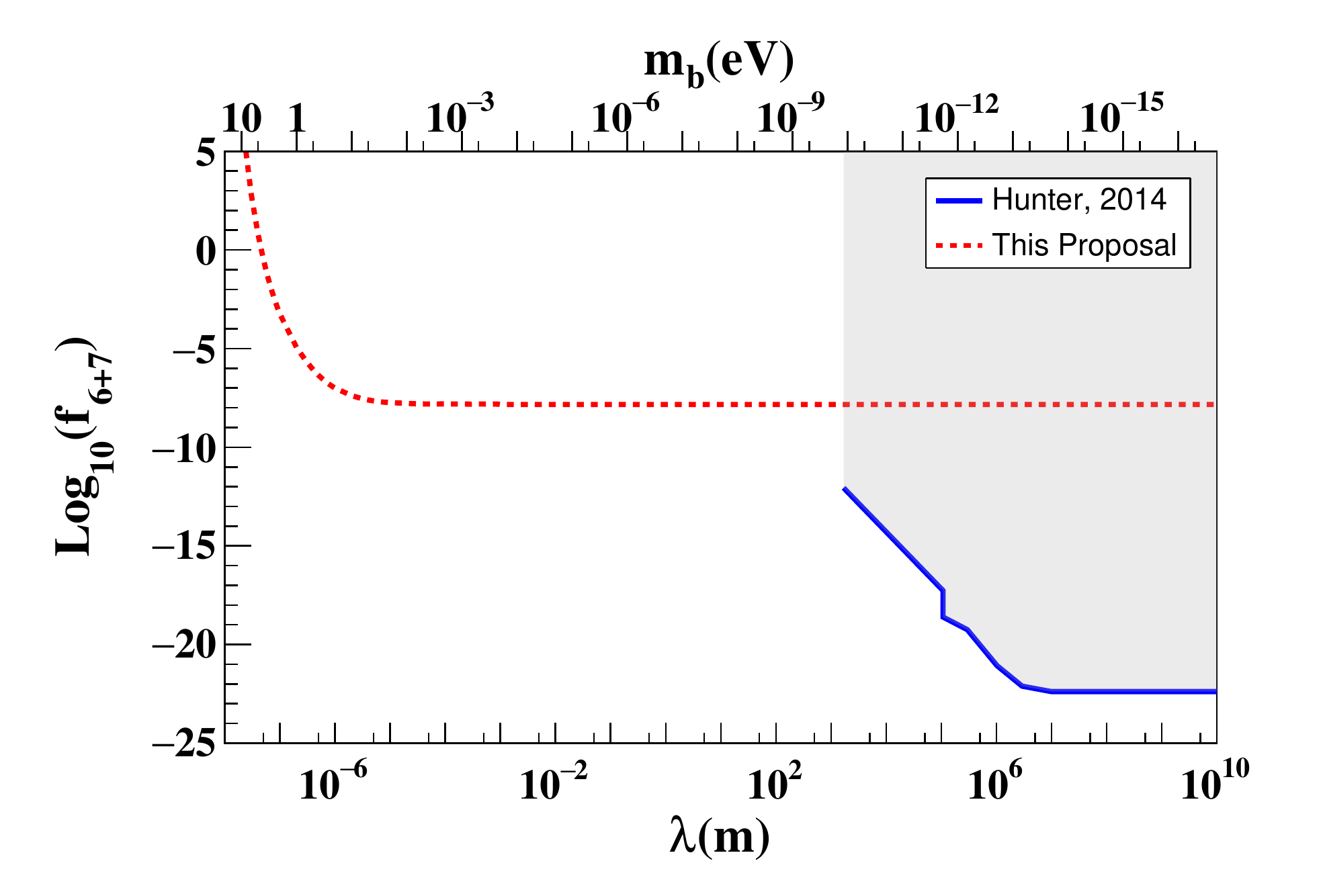}
\caption{\textbf{Exclusion plot for electron-nuclear interaction $\bm{V_{6+7}}$.} Exclusion region of $f_{6+7}$ for electron-neutron coupling by a previous experiment: Hunter, 2014~\cite{Hunter:2014}. The minimum detectable electron-nuclear $f_{6+7}$ for our proposed experiment is shown in dashed red. }
\label{fig:sensitivity_6}
\end{figure}

Figure~\ref{fig:sensitivity_6} shows the calculated constraints on the coupling constant $f_{6+7}$ for electron-nuclear spin coupling. To our knowledge, the only other experiment probing $f_{6+7}$ for electron-nuclear spin coupling used a $^{199}$Hg-Cs co-magnetometer~\cite{Hunter:2014} to set constraints at length scales $\gtrsim1~{\rm km}$. Another experiment used a K-Rb atomic co-magnetometer with SmCo$_5$~\cite{Ji:2018tny} to set limits on $f_{6+7}$ at the $\sim10~{\rm cm}$ length scale, but in this case both spin species were electron spins. Other technologies have also been proposed to set limits on $f_{6+7}$ for electron-electron coupling at the centimeter~\cite{Chu:2016} and $\gtrsim100~{\rm \upmu m}$~\cite{Leslie:2014} length scales. In any case, our proposed experiments could provide the first constraints on $f_{6+7}$ for interaction lengths in the $1\mbox{-}100~{\rm \upmu m}$ range.

\subsection{Interaction \texorpdfstring{$\bm{V_{14}}$}{}}
\label{sec:v14}

The interaction $V_{14}$ can be mediated by spin-1 bosons~\cite{Dobrescu:2006,Leslie:2014}. The potential is defined as:
\begin{align}
V_{14} & = f_{14}\frac{\hbar}{4\pi} [\bm{\hat{\sigma}_{\rm nv}}\cdot(\bm{\hat{\sigma}_{\rm tm}}\times\bm{v})]\left(\frac{1}{r}\right)e^{-r/\lambda}.
\label{eq:v14}
\end{align}
The coupling strength $f_{14}$ is defined as the product of the axial vector coupling of the NV center electron ($g_A^{e}$) and the axial vector coupling of the test mass polarized nucleon ($g_A^{N}$), $f_{14}\equiv g_A^{e} g_A^{N}$~\cite{Leslie:2014}. The effective magnetic field due to $V_{14}$ is: 
\begin{align}
B_{14} & = f_{14}\frac{\rho}{2\pi\gamma} \int_{mass}[\bm{\hat{\sigma}_{\rm nv}}\cdot(\bm{\hat{\sigma}_{\rm tm}}\times\bm{v})]\left(\frac{1}{r}\right)e^{-r/\lambda}d^3r_{\rm tm}.
\label{eq:b14}
\end{align}
The experimental geometry to detect $V_{14}$ is shown in Fig.~\ref{fig:angle_adjust}(c). As for potential $V_{6+7}$, the spin direction of the polarized mass, $\bm{\hat{\sigma}_{\rm tm}}$, is oriented at an angle $\theta\approx80\degree$ with respect to the NV spin direction, $\bm{\hat{\sigma}_{\rm nv}}$, Secs.~\ref{sec:pulsepol},\ref{sec:gradients}. In contrast to $V_{6+7}$, the movement of the polarized test mass is perpendicular to $\bm{\hat{\sigma}_{\rm nv}}$. 

 \begin{figure}[htb]
\includegraphics[width=\columnwidth]{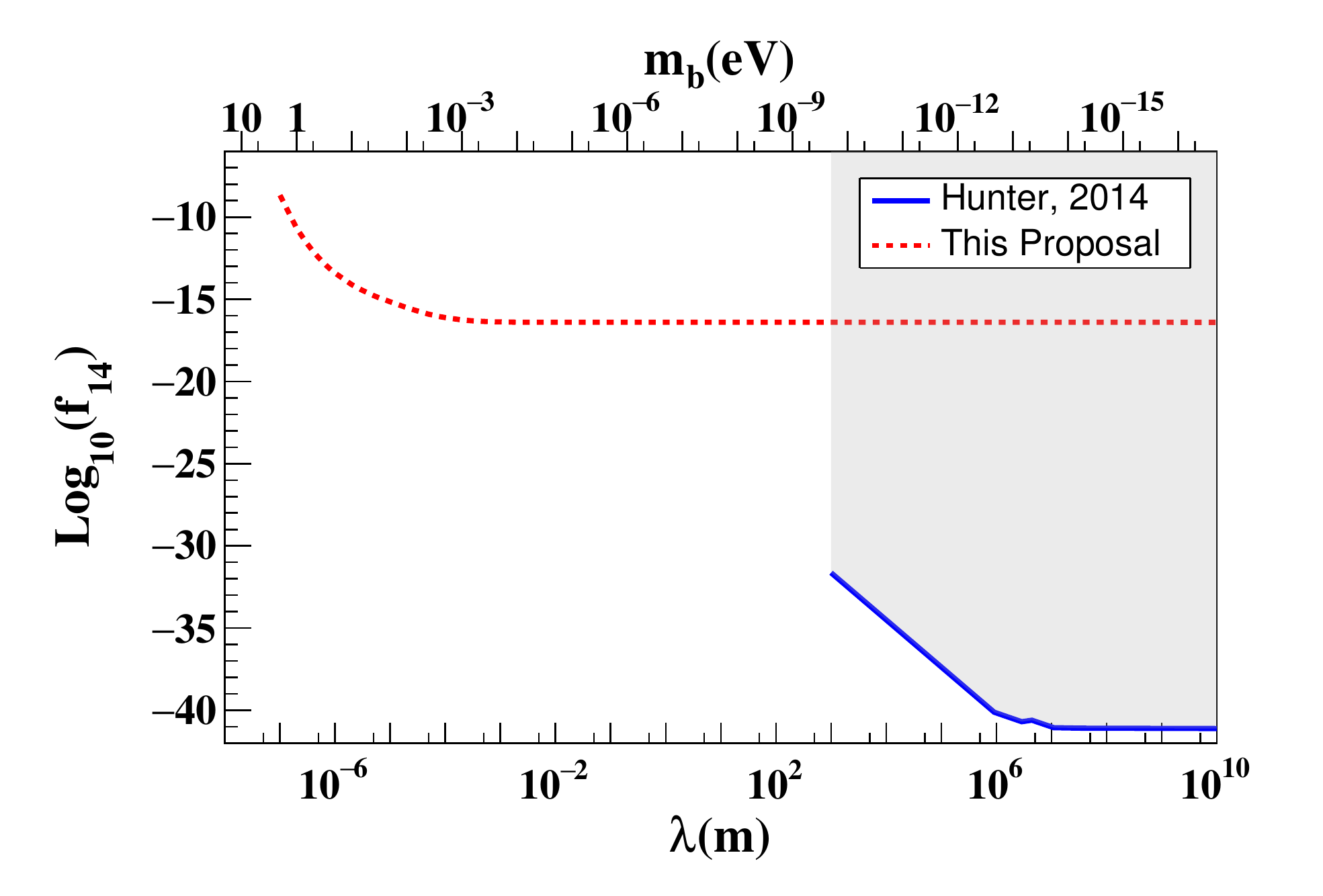}
\caption{\textbf{Exclusion plot for electron-nuclear interaction $\bm{V_{14}}$.} Exclusion region of $f_{14}$ for electron-neutron spin coupling by a previous experiment: Hunter, 2014~\cite{Hunter:2014}. The minimum detectable electron-nuclear $f_{14}$ for our proposed experiment is shown in dashed red.}
\label{fig:sensitivity_14}
\end{figure}

 Figure~\ref{fig:sensitivity_14} shows the calculated constraints for the coupling constant $f_{14}$. The only prior experimental constraint on $f_{14}$ that we are aware of was set with a $^{199}$Hg-Cs comagnetometer~\cite{Hunter:2014}, but only at length scales $\lambda\gtrsim1~{\rm km}$.
There is no constraint in the small interaction length range, but several proposed experiments including diamond magnetometers, atomic magnetometers~\cite{Chu:2016}, and the torsional oscillator~\cite{Leslie:2014}, may have the potential to explore it.

\subsection{Interaction \texorpdfstring{$\bm{V_{15}}$}{}}
\label{sec:v15}

The interaction $V_{15}$ can be mediated by spin-1 bosons~\cite{Dobrescu:2006,Leslie:2014}. The potential is defined as
\begin{align}
V_{15} & = -f_{15}\frac{\hbar^3}{8\pi m_{e} m_{N} c^2 }\nonumber\\
&\times\{[\bm{\hat{\sigma}_{\rm nv}}\cdot(\bm{v}\times\bm{\hat{r}})](\bm{\hat{\sigma}_{\rm tm}}\cdot\bm{\hat{r}})+(\bm{\hat{\sigma}_{\rm nv}}\cdot\bm{\hat{r}})[\bm{\hat{\sigma}_{\rm tm}}\cdot(\bm{v}\times\bm{\hat{r}})]\}\nonumber\\
&\times\left(\frac{1}{\lambda^2 r}+\frac{3}{\lambda r^2}+\frac{3}{r^3}\right)e^{-r/\lambda},
\label{eq:v15}
\end{align}
The coupling strength $f_{15}$ is defined as the product of the vector coupling of the NV center electron ($g_V^{e}$) and the vector coupling of test mass polarized nucleon ($g_V^{N}$), $f_{15}=g_V^{e} g_V^{N}$~\cite{Leslie:2014}. 

The effective magnetic field due to $V_{15}$ is:
\begin{align}
B_{15} & = -f_{15}\frac{\rho\hbar^2}{4\pi\gamma m_{e} m_{N} c^2 }\nonumber\\
&\times\int\{[\bm{\hat{\sigma}_{\rm nv}}\cdot(\bm{v}\times\bm{\hat{r}})](\bm{\hat{\sigma}_{\rm tm}}\cdot\bm{\hat{r}})+(\bm{\hat{\sigma}_{\rm nv}}\cdot\bm{\hat{r}})[\bm{\hat{\sigma}_{\rm tm}}\cdot(\bm{v}\times\bm{\hat{r}})]\}\nonumber\\
&\times\left(\frac{1}{\lambda^2 r}+\frac{3}{\lambda r^2}+\frac{3}{r^3}\right)e^{-r/\lambda}d^3r_{\rm tm},
\label{eq:b15}
\end{align}

\begin{figure}[htb]
\includegraphics[width=\textwidth]{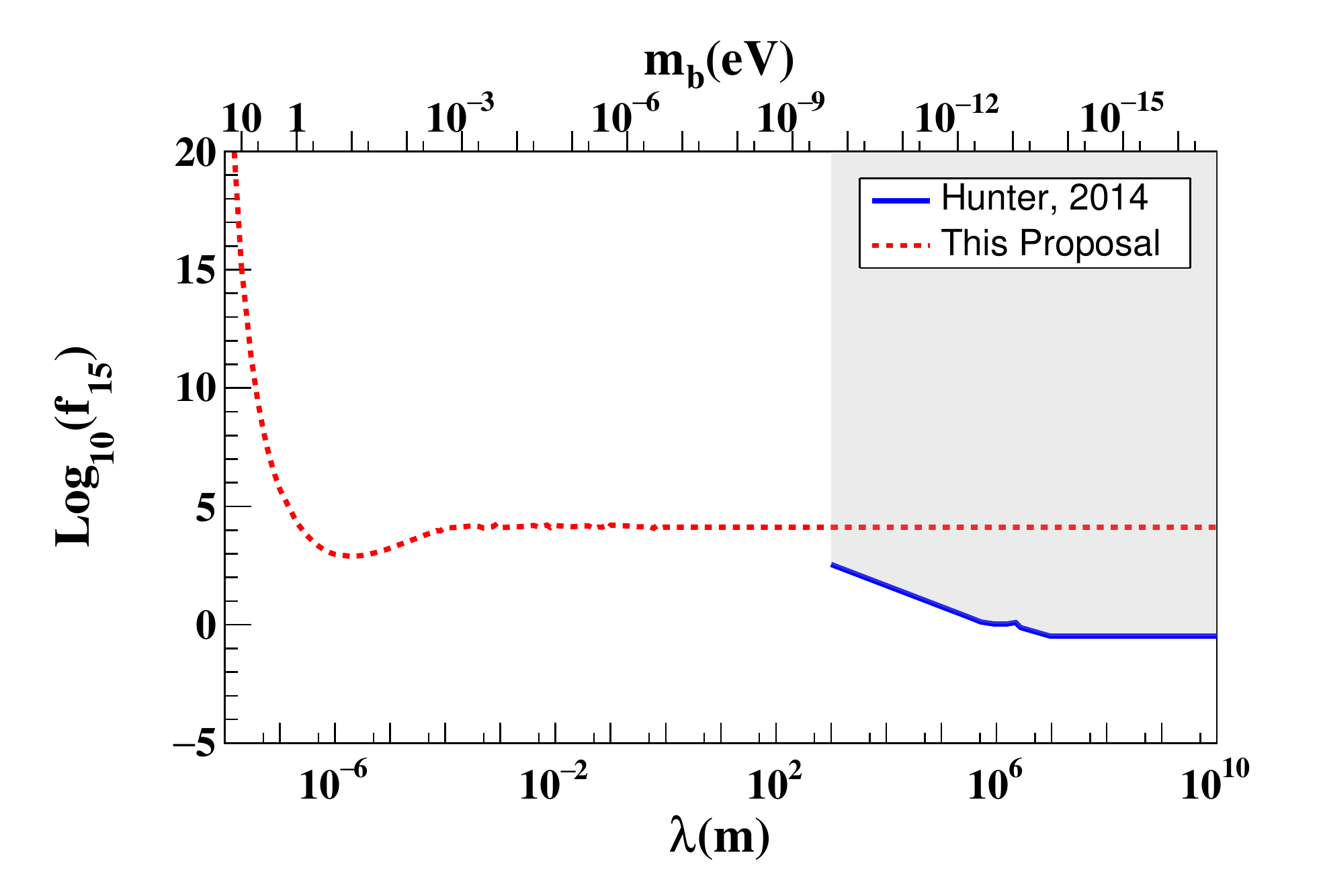}
\caption{\textbf{Exclusion plot for electron-nuclear interaction $\bm{V_{15}}$.} Exclusion region of $f_{15}$ for electron-neutron coupling by a previous experiment: Hunter, 2014~\cite{Hunter:2014}. The minimum detectable electron-nuclear $f_{15}$ for our proposed experiment is shown in dashed red.}
\label{fig:sensitivity_15}
\end{figure}

As shown in Fig.~\ref{fig:angle_adjust}(c), the experimental geometry for $V_{15}$ is identical to that of $V_{14}$. Figure~\ref{fig:sensitivity_15} shows the calculated constraints of our proposed experiments for the coupling constant $f_{15}$. As is the case for $f_{6+7}$ and $f_{14}$, the only prior experimental constraint on $f_{15}$ that we are aware of was set with a $^{199}$Hg-Cs comagnetometer~\cite{Hunter:2014}, but only at length scales $\lambda\gtrsim1~{\rm km}$.
There are tight constraints on the electron-electron $f_{15}$ coupling from the dark photon limit in Ref.~\cite{Jaeckel:2010,Essig:2013lka}, but not for electron-nuclear coupling.

\section{Test mass considerations}
\label{subsec:testmass}
\subsection{Unpolarized test mass}
There are a number of materials to choose from for the unpolarized test masses used to probe $V_{4+5}$ and $V_{12+13}$. $\text{SiO}_2$ has been used in Refs.~\cite{Rong:2017,rong:2020observation} with a nucleon density of $1.33\times 10^{30}~{\rm m^{-3}}$. Bismuth germanate (BGO) has an even higher nucleon density ($4.29\times 10^{30}~{\rm m^{-3}}$) and has been used in experiments probing larger interactions lengths~\cite{Tullney:2013}. High purity single crystals of different dimensions are commercially available and thin films of BGO can also be deposited for the thinner experimental geometries. We selected $\text{SiO}_2$ for calculations because of its ease of integration in mechanical oscillators, Sec.~\ref{sec:mechanics}.

\subsection{Polarized test mass}
For potentials $V_{6+7}$, $V_{14}$, and $V_{15}$, a high polarized spin density is desirable. In addition, using low gyromagnetic-ratio spins is desirable in order to minimize magnetic portion of the sensor-test mass coupling. Previous proposals addressed this challenge for detecting electron-electron interactions\cite{Leslie:2014}. However, to probe electron-nuclear spin interactions, a different solution is needed.

\begin{table}[ht]
\centering
\begin{tabularx}{\columnwidth}{|C|C|C|}
\hline
\textbf{Material} & \textbf{Polarized nuclear spin density, $\rho$ (m$^{-3}$)} & \textbf{Magnetization (A/m)}\\
\hline\hline
$^{29}$Si ($100\%$)& $8 \times 10^{19}$ & $2 \times 10^{-7}$ \\ \hline
BGO & $7\times 10^{19}$ & $1.4\times 10^{-7}$\\ \hline
NaI & $5\times10^{23}$ & $1\times10^{-3}$ \\ \hline
Diamond $1.1\%$ $^{13}$C & $5\times 10^{25}$ &$0.02$ \\ \hline
\end{tabularx}
\caption{\textbf{polarized test mass materials}. Estimates for the polarized spin density ($\rho$) and the corresponding magnetization are provided for four choices of test mass material. The values for $^{29}$Si, BGO, and NAI assume thermal equilibrium at room temperature ($300~{\rm K}$) and $10~{\rm mT}$ bias field. The value for diamond assumes a $^{13}$C hyperpolarization level similar to that observed experimentally in Ref.~\cite{Alvarez:2015}}
\label{tab:polarized_mass}
\end{table}

Realizing a high nuclear-spin polarization at room temperature is challenging considering the low magnetic fields considered in this work. At room temperature and $10~{\rm mT}$, the excess polarized nuclear spin density of Silicon and BGO is $8\times10^{19}~{\rm m^{-3}}$ and $7\times10^{19}~{\rm m^{-3}}$, respectively, Tab.~\ref{tab:polarized_mass}. Materials with large nuclear quadrupole splittings can enable larger spin polarizations. Quadrupole splittings of 200 MHz and higher are commonly reported for materials containing heavy atoms such as iodine~\cite{ludwig_bromine_1956,miller_hyperfine_1969}. Sodium iodide is a popular scintillator material and high purity single-crystal samples are commercially available, which makes it an appealing candidate material. However, the source of polarization is still thermal and at room temperature, the excess polarized nuclear spin density is still only approximately $5\times10^{23}~{\rm m^{-3}}$, Tab.~\ref{tab:polarized_mass}.

\subsection{Nuclear-spin-polarized diamond test mass}
\label{sec:pulsepol}
A potent way of increasing the polarized nuclear spin density in the test mass is to use nuclear spins with a polarization exceeding the thermal equilibrium (``hyperpolarized''). The nuclear-spin-polarized test mass we analyze in this work uses optically-hyperpolarized $^{13}$C nuclei (natural abundance of $1.1\%$) in diamond. The hyperpolarization mechanism relies on the transfer of optical polarization from NV electron spins to $^{13}$C nuclear spins in the diamond lattice, and it works at room temperature and a wide range of magnetic fields.  To realize the polarization transfer, various methods have been explored including direct flip-flop~\cite{Pagliero:2018, pagliero_optically_2020}, continuous microwave drive \cite{Cai:2013, Hovav:2018, London:2013, King:2015, Alvarez:2015}, frequency-swept microwave drive \cite{Ajoy:2018, Ajoy:2018:enhanced, Zangara:2019}, and pulsed microwave techniques \cite{Schwartz:2018, Broadway:2018}. With these methods, polarized spin densities up to $5\times10^{25}~{\rm spins/cm}^3$~\cite{Alvarez:2015} have been experimentally realized. 

Another convenient option could be to hyperpolarize the native nitrogen nuclear spins associated with NV centers in diamond. This could provide the intriguing possibility to detect nuclear-nuclear interactions by performing quantum sensing on the nitrogen nuclear spins~\cite{Jarmola:2020,Jarmola:2021}. However, here we focus on $^{13}$C nuclei owing to their high density.

\begin{figure}
    \centering
    \includegraphics[width=0.7\columnwidth]{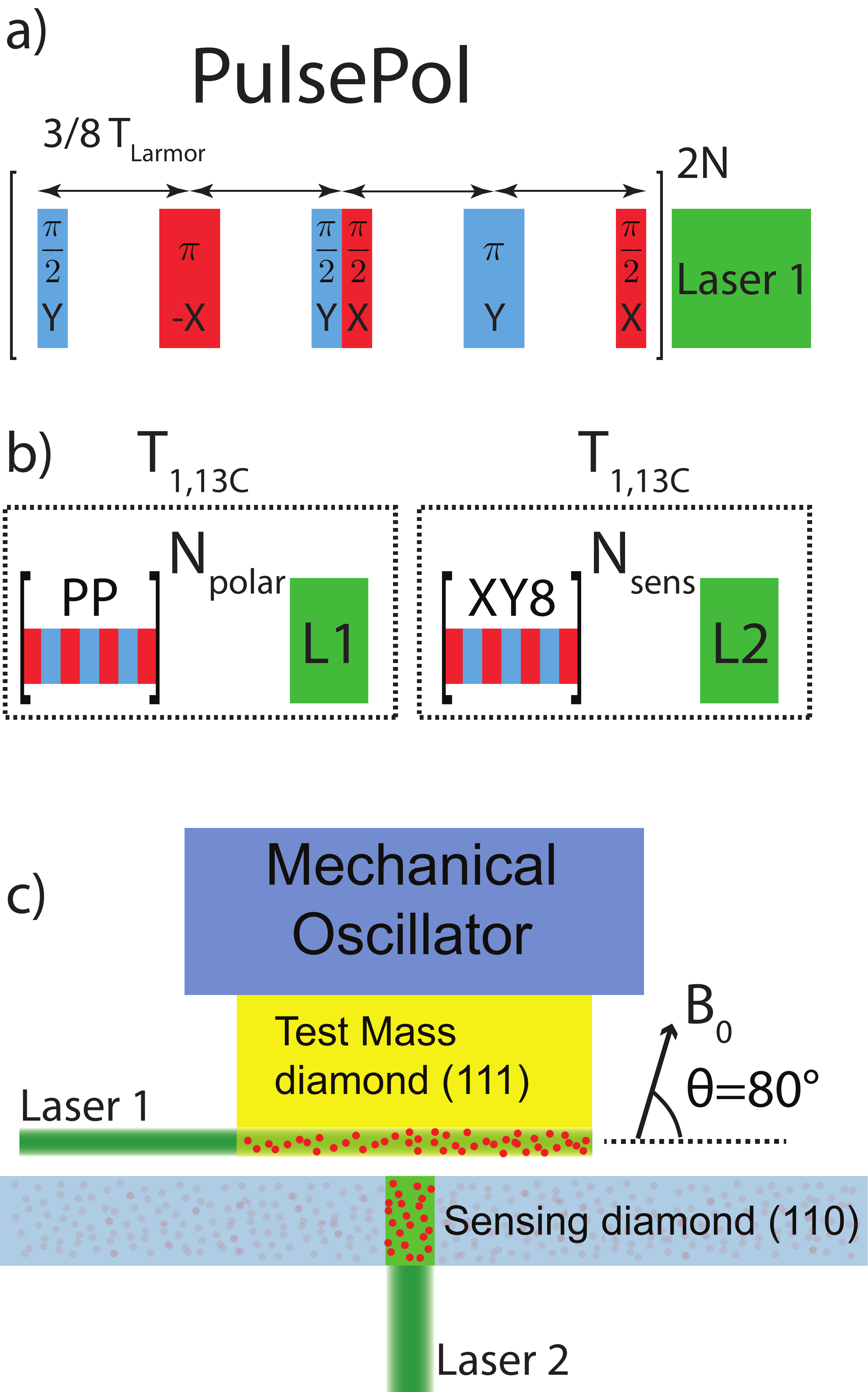}
    \caption{\textbf{Pulse sequence for hyperpolarized diamond test mass.} a) The PulsePol sequence. The spacing between pulses is determined by the $^{13}$C Larmor period, $T_{\rm Larmor}$. The sequence is repeated for a time comparable to the NV coherence time, $T_{\rm 2,NV}$. b) Complete pulse sequence for probing interactions with hyperpolarized diamond test masses. The full measurement sequence consists of repeating PulsePol pulses for $T_{\rm 1,^{13}C}\approx10~{\rm s}$ followed by a readout sequence [Fig.~\ref{fig:pulse}(c)] of the same duration. c) Geometry for probing interactions with a hyperpolarized diamond test mass. The test mass and sensor diamonds are cut with a different surface polish and are addressed with separate laser beams. Thin ($\lesssim100~{\rm nm}$) reflective coatings may be used to minimize cross-talk from stray laser light.}
    \label{fig:pulsepol}
\end{figure}

For a single-crystal diamond test mass at a field of $\sim 10~{\rm mT}$, the PulsePol hyperpolarization protocol \cite{Schwartz:2018} is promising for its relatively high transfer efficiency and robustness against microwave errors. The pulse sequence is depicted in Fig.~\ref{fig:pulsepol}(a,b). The effective Hamiltonian governing the time-evolution of the system is $H_{\text{eff}} = A_{x}I_x S_x$, where $A_x$ is the transverse hyperfine coupling constant between the spin of the NV center and the carbon nucleus and $I_x$ and $S_x$ are nuclear and electron spin operators respectively. The effective Hamiltonian leads to flip-flops between the NV center and the target nucleus, which ultimately transfers polarization from the former to the latter. The duration of each PulsePol sequence is determined by the NV spin coherence time of the NV centers, $T_{\rm 2,NV}$. After each PulsePol cycle, the NV centers are optically repolarized and the whole process is repeated for a time equal to the spin relaxation time of the $^{13}$C nuclei, $T_{\rm 1,^{13}C}\approx10~{\rm s}$~ \cite{Henshaw18334}). After the hyperpolarization sequence, the detection sequence shown in Fig.~\ref{fig:pulse}(c) is used to probe the exotic interaction for a time $T_{\rm 1,^{13}C}$, after which the nuclei have to be repolarized and the entire process repeats. 

In order to maximize the interaction strength for potentials $V_{6+7}$ and $V_{14},V_{15}$, the angle between the test mass nuclear spin polarization, $\bm{\hat{\sigma}_{\rm tm}}$, and the sensor NV electron spin polarization, $\bm{\hat{\sigma}_{\rm nv}}$, should be close to $90\degree$, Eqs.~\eqref{eq:pot_v67},\eqref{eq:v14},\eqref{eq:v15}. Under the PulsePol sequence, $\bm{\hat{\sigma}_{\rm tm}}$ tends to align with the axis of the NV centers that are on resonance with the microwaves. This implies that the NV axis addressed in the test mass should be orthogonal to the NV axis addressed in the sensor. This can be realized by using a (110) surface polish for the sensing diamond (NV axis in plane) and a (111) surface polish for the test mass diamond (NV axis normal to the surface), Fig.~\ref{fig:pulsepol}(c). 

\begin{figure}
    \centering
    \includegraphics[width=\columnwidth]{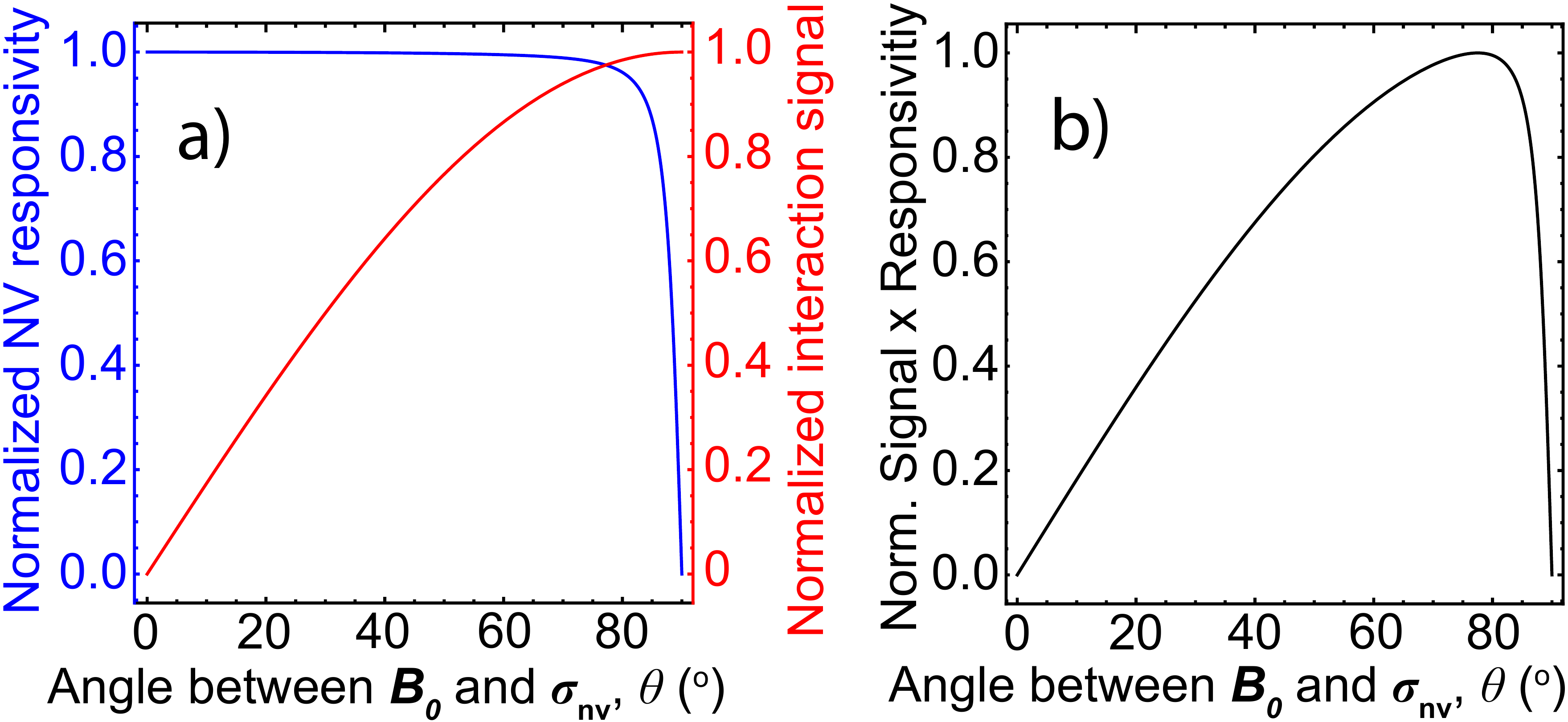}
    \caption{\textbf{Choice of field angle $\bm{\theta}$ for polarized test mass}. a) The NV sensor responsivity (blue) and the  interaction signal strength (red) as a function of the angle $\theta$ between the bias field, $B_0$ and the sensor NV polarization axis $\bm{\hat{\sigma}_{\rm nv}}$ (see Fig.~\ref{fig:angle_adjust}). The responsivity is computed assuming a bias field strength of $10~{\rm mT}$. b) The normalized signal-weighted responsivity as a function of $\theta$. Optimum sensing condition are reached when this value is maximized, at $\theta\approx80\degree$. }
    \label{fig:responsivity_v_signal}
\end{figure}

A remaining consideration is which angle $\theta$ the bias magnetic field should be applied with respect to $\bm{\hat{\sigma}_{\rm nv}}$. On one hand, the nuclear spin polarization, and thus the exotic interaction signal strength, is maximized when the field is along $\bm{\hat{\sigma}_{\rm tm}}$, $\theta=90\degree$, Fig.~\ref{fig:responsivity_v_signal}(a). This is because only the component of $\bm{\hat{\sigma}_{\rm tm}}$ along the applied magnetic field will be conserved (transverse components will decay on a much faster timescale than $T_{\rm 1,^{13}C}$). On the other hand, the responsivity of the NV sensor (the change in spin transition frequency for a given interaction field strength), and thus the sensitivity, approaches zero when $\theta=90\degree$, Fig.~\ref{fig:responsivity_v_signal}(a). The figure of merit is the product of the two values, which reaches maximum at $\theta\approx80\degree$ for a bias field of $10~{\rm mT}$, Fig.~\ref{fig:responsivity_v_signal}(b).

\section{Driving Test mass Movement}
\label{sec:mechanics}
In order to move the test mass, we seek a mechanical oscillator that can generate movement in the lateral (in-plane) mode, with nano-scale resolution, ${\sim}1~{\rm \upmu m}$ range, and high peak velocity. To accomplish this, we consider using microelectromechanical system (MEMS) oscillators with two different transduction mechanisms: electrostatic comb drives and piezoelectric actuation. We estimate that a displacement amplitude of $d_1\approx0.75~{\rm \upmu m}$ ($1.5~{\rm \upmu m}$ peak-peak) and a frequency $f_m\approx1~{\rm MHz}$ should be possible, using these mechanisms. 

Previous MEMS electrostatic comb drive systems for controlling lateral position have demonstrated displacements $>28~{\rm \upmu m}$, with sub-micron resolution, but most of these devices have low operating frequencies in the $0.1\mbox{-}10~{\rm kHz}$ range~\cite{LASZCZYK2010255}. Attempts to create high-velocity lateral actuation MEMS devices have realized velocities of approximately $1.5~{\rm m/s}$~\cite{Eltagoury:2016}. The three commonly used transduction mechanisms for nanopositioners are based on electrostatic, electro-thermal, or piezoelectric actuation. Electrothermal actuators tend to be slow and require high power and thus are not suitable for this application. Piezoelectrics have high operating frequencies with nano-scale resolution, but they are typically designed for out-of-plane displacement. In-plane displacement is possible with piezoelectrics using double-ended tuning fork resonator designs. They can achieve high frequencies ($\gtrsim1~{\rm MHz}$), but with low displacement ($\ll1~{\rm \upmu m}$)~\cite{Olsson:2009,Wang:2017}. Electrostatic comb drives offer a relatively fast response time and precise lateral displacement with relatively high force, and thus they appear most promising for our application. However, previous comb drive designs often operated at lower frequencies, so a new design is needed to increase the velocity.

\begin{figure}[htb]
    \centering
    \includegraphics[width=\columnwidth]{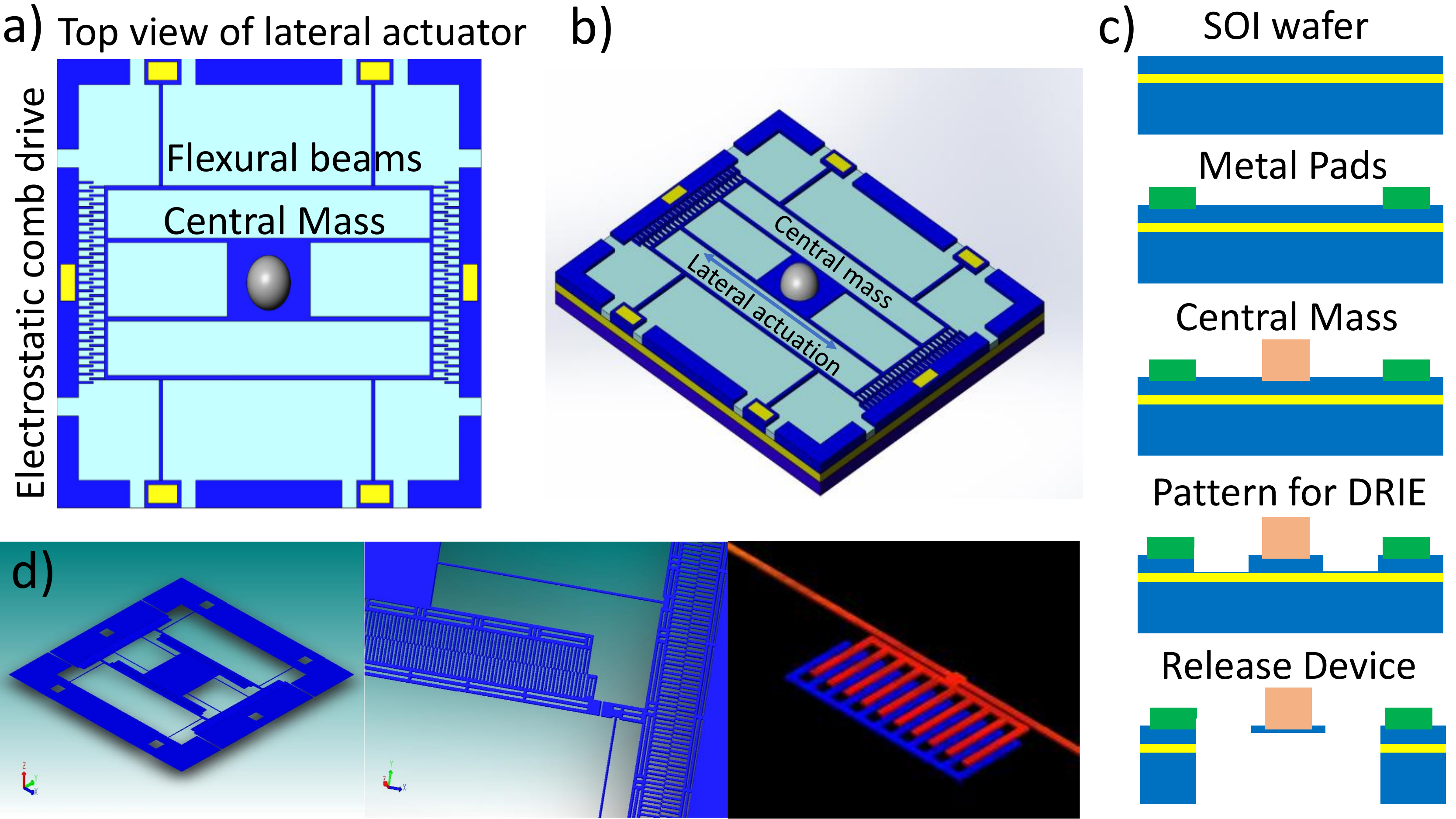}
    \caption{\textbf{MEMS mechanical oscillator design.} a) Top view of the design of a lateral silicon mechanical oscillator, with test mass. It is actuated by an interdigitated (IDT) comb. b) Side view of the device. c) Cross section of the process flow for fabricating the device using Silicon-on-Insulator (SOI) wafer. DRIE--deep reactive ion etching. d) Images from COMSOL simulation of the resonator device.}
    \label{fig:comb_drive_design}
\end{figure}
    
Our proposed high-frequency comb-drive design, Fig. \ref{fig:comb_drive_design}, consists of three main components i)interdigitated (IDT) electrode fingers (``comb drives''), to create the actuation, ii) a shuttle for the central movable mass, and iii) folded flexural beams for mechanical compliance. The device consists of multiple IDTs, where each group consists of one set of IDTs that are anchored (stationary) and another set that is free to move via flexural beams. An AC voltage applied between the two sets of fingers causes a lateral displacement by varying the capacitance between the IDTs. The free-standing structures have low-stiffness flexural arms and a central thin-film plate where the test mass is attached. The flexural arms are designed to allow the thin film plate to move laterally with high accuracy and nano-scale resolution. High stiffness beams are used to control the direction of actuation.

The frequency and velocity that can be achieved by the MEMS device is determined by the stiffness of the beams, the overall mass of the system, and damping factors. A stiffer beam with lower mass will enable us to realize resonant frequencies approaching $1~{\rm MHz}$. The frequency can be further increased by i) reducing the number of folds in the flexural beams, ii) optimizing the flexural beam dimensions to increase stiffness (reduce length, increase thickness, and increase width), iii) reduce the thickness of the central plate mass, and iv) using a high elastic modulus material such as (111) Si. The comb-drive can be fabricated from Silicon-on-Insulator (SOI) wafers with (111) Si using deep reactive ion etching, where the thickness of the IDT and flexural beams determined the device layer thickness of the SOI wafer. The force can be optimized based on the pitch of the IDT and the number of fingers. The peak displacement is dependent on the applied voltage and can be controlled with nano-scale resolution and micro-scale range.

\section{Potential sources of systematic error}
\label{sec:systematics}
To reliably establish new constraints on exotic spin-dependent interactions it is important to exclude the coupling of the NV electron spin to its environment through any "non-exotic" interactions or "systematic errors". The phase-sensitive nature of our measurement protocol, Fig.~\ref{fig:pulse}(c), renders it highly frequency selective, with a resolution $\ll1~{\rm Hz}$ that is only limited by the clock of the control electronics or averaging time. This means that we must only concern ourselves with signals that coherently oscillate at $f_m\approx1~{\rm MHz}$ and couple to the NV spin system or those that generate a significant background noise that reduces overall sensitivity. For example, experimental ``drifts'' (such as changes in the laser intensity, microwave phase or amplitude, or NV spin transition frequencies) may limit the realizable signal averaging time if they are not stabilized. However they are not expected to produce a false signal, so they are not discussed here.

\subsection{Air friction and shear stress} 
 If the experiment is performed in atmospheric conditions, the friction of the air layer between the test mass and sensing diamond should be considered. Assuming a no-slip boundary between the air and the test-mass and sensor surfaces, the movement of the test mass leads to a shear stress in the sensing diamond. Shear stress, in turn, can produce shifts in the NV spin transition frequencies. Only the stress component along the NV axis leads to appreciable shifts, so this effect is a potential systematic error for the geometry used to probe potentials $V_{12+13}$, $V_{14},$ and $V_{15}$.
 
 The shear stress experienced by the sensing diamond can be written as $\tau_{\rm stress} \approx \mu v/d_{\rm gap}$, where $\tau_{\rm stress}$ is the shear stress, $\mu=1.8\times10^{-5}~{\rm Pa\cdot s}$ is the dynamic viscosity of air, $v$ is the velocity of the test mass, and $d_{\rm gap}$ is the distance between the test mass and the sensor. The largest shift occurs for $V_{15}$ at target $\lambda = 0.5~{\rm \upmu m}$. For $v=4.7~{\rm m/s}$ and $d_{\rm gap}=0.2~{\rm \upmu m}$, the shear stress component along the NV axis is $\tau_{\text{stress}} \approx 400~{\rm Pa}$. With an on-axis stress-spin coupling constant of $21~{\rm MHz/GPa}$ \cite{PhysRevB.100.174103,PhysRevB.99.174102}, this would lead to shifts of the NV spin transition frequencies of ${\sim}~8~{\rm Hz}$ or an equivalent magnetic field, when monitoring only a single NV transition, of ${\sim}300~{\rm pT}$. 
 
 These frequency shifts are more than an order of magnitude larger than the shot-noise limited sensitivity and must be mitigated in some manner. One option is to perform the experiment under vacuum ($10^{-4}~{\rm bar}$ or lower). A more convenient options is to exploit the fact that the energy shift due to shear stress is the same for both $m_s=0\leftrightarrow m_s\pm1$ transitions, while the shift due to the exotic interactions inverts its sign under the change of transition. By alternately addressing both NV spin transitions, it is thus possible to suppress the shift due to shear stress.
 
 \subsection{Stark shifts due to surface charges}
 We next consider the effect of AC electric fields coupling to the sensing spin through the Stark shift. The NV spin transition frequencies, $f_{\pm}$, shift with electric field, $\bm{E}$, as $f_{\pm}\approx d_{\|}E_{\|}\pm\sqrt{(\gamma_{\rm nv}B_{\|})^2+(d_{\perp}E_{\perp})^2}$, where the $_{\|}$ and $_{\perp}$ directions are with respect to the NV axis ($\bm{\hat{\sigma}_{\rm nv}}$) and the NV electric dipole coupling parameters are $d_{\|}\approx3.5~{\rm kHz/(MV/m)}$ and $d_{\perp}\approx170~{\rm kHz/(MV/m)}$~\cite{VANOORT1990}.
 
 Surface charges on the surface of the test mass could produce a substantial electric field normal to the surface which, for every geometry considered here, would point orthogonal to $\bm{\hat{\sigma}_{\rm nv}}$. However, if the spacing between surface charges is much larger than $d_{\rm gap}$, the Stark shifts would be mostly constant with the motion, since the radial dimension of the test mass is much larger than the displacement, $R_{\rm tm}\gg d_1$. Furthermore, any residual electric field would likely be in phase with the displacement trajectory of the test mass but $90\degree$ out of phase with the velocity modulation. Since the interaction effective fields are in phase with the velocity modulation, the phase-sensitive nature of our detection protocol would further suppress the effect of such electric fields. 
 
For completeness, let's suppose that, due to misalignments, the test mass motion produces an AC electric field of $E_{\perp}=1~{\rm kV/m}$ that is in phase with the velocity. For the geometries with unpolarized test mass, where $B_0\approx10~{\rm mT}$ is applied along $\bm{\hat{\sigma}_{\rm nv}}$, the NV frequencies would experience shifts $f_{\pm}\approx\pm50~{\rm \upmu Hz}$ which would mimic an AC magnetic field ${\sim}2~{\rm fT}$. For the geometries with polarized test mass, where $B_0\approx10~{\rm mT}$ is applied at $\theta\approx80\degree$ with respect to $\bm{\hat{\sigma}_{\rm nv}}$, the shifts would be $f_{\pm}\approx\pm0.3~{\rm mHz}$ which would mimic an AC magnetic field ${\sim}10~{\rm fT}$. This is smaller than the minimum detectable field of the diamond sensor for all geometries considered here, $\delta B_{\rm min}=10\mbox{-}100~{\rm fT}$ after $t=10^4~{\rm s}$ of averaging. The effect of any residual AC electric fields $E_{\|}$ along the NV axis would be even smaller. Moreover, any residual shift due to $E_{\|}$ would be the same for both $f_{\pm}$ transitions and can be eliminated by alternately probing each NV spin transitions. Thus, this effect is expected to be negligible for all the potentials considered here.

\subsection{Magnetic field from moving surface charges}
A moving test mass with surface charge constitutes a planar current which produces a magnetic field, transverse to the axis of motion. This effect would impact experiments that probe potentials $V_{4+5}$, $V_{14}$, and $V_{15}$. 

First, we consider the worst-case scenario--the surface charge density produces an electric field at the threshold of inducing dielectric breakdown in air ($E\approx3~{\rm MV/m}$). Since, $R_{\rm tm}^2\gg A_{\rm nv}$, we can approximate the test mass surface charges as an infinite sheet. The electric field is given by $E=\sigma/(2\epsilon_0)$, where $\sigma$ is the surface charge density and $\epsilon_0$ is the vacuum dielectric constant. Setting $E=3~{\rm MV/m}$, we arrive at a surface charge density $\sigma\approx50~{\rm \upmu C/m^2}$. The peak surface current density is $J= \sigma v \approx2.5~{\rm mA/m}$, resulting in a magnetic field $B=\mu_0 J/2\approx150~{\rm pT}$, where $\mu_0$ is the vacuum permeability. This spurious field is orthogonal to the NV spin axis for $V_{12+13}$ and $V_{6+7}$ and can be neglected for those interactions. However for the other potentials, the systematic error is several orders of magnitude larger than $\delta B_{\rm min}\approx1~{\rm pT}$ and must be addressed.

One possible solution to reduce the induced magnetic field is to coat the test mass with a thin layer of low affinity triboelectric material. A silicone rubber such as Powersil has been reported to have a maximum surface charge density of $30~{\rm \upmu C/m^2}$~\cite{Kumara:2011}, which would correspond to a $100~{\rm pT}$ magnetic field at the sensing spin. Other measures such as enhanced gas neutralization can be used to reduce the surface charge further~\cite{Liu:2020}. If we assume a surface charge density of $10~{\rm \upmu C/m^2}$ from a low affinity triboelectric coating, then the spurious field is $B\approx30~{\rm pT}$. Another possible solution would be decreasing the test mass radius $R_{\rm tm}$ or increasing the test mass-sensor gap $d_{\rm gap}$.

\begin{figure}[htb]
    \centering
    \includegraphics[width=1\textwidth]{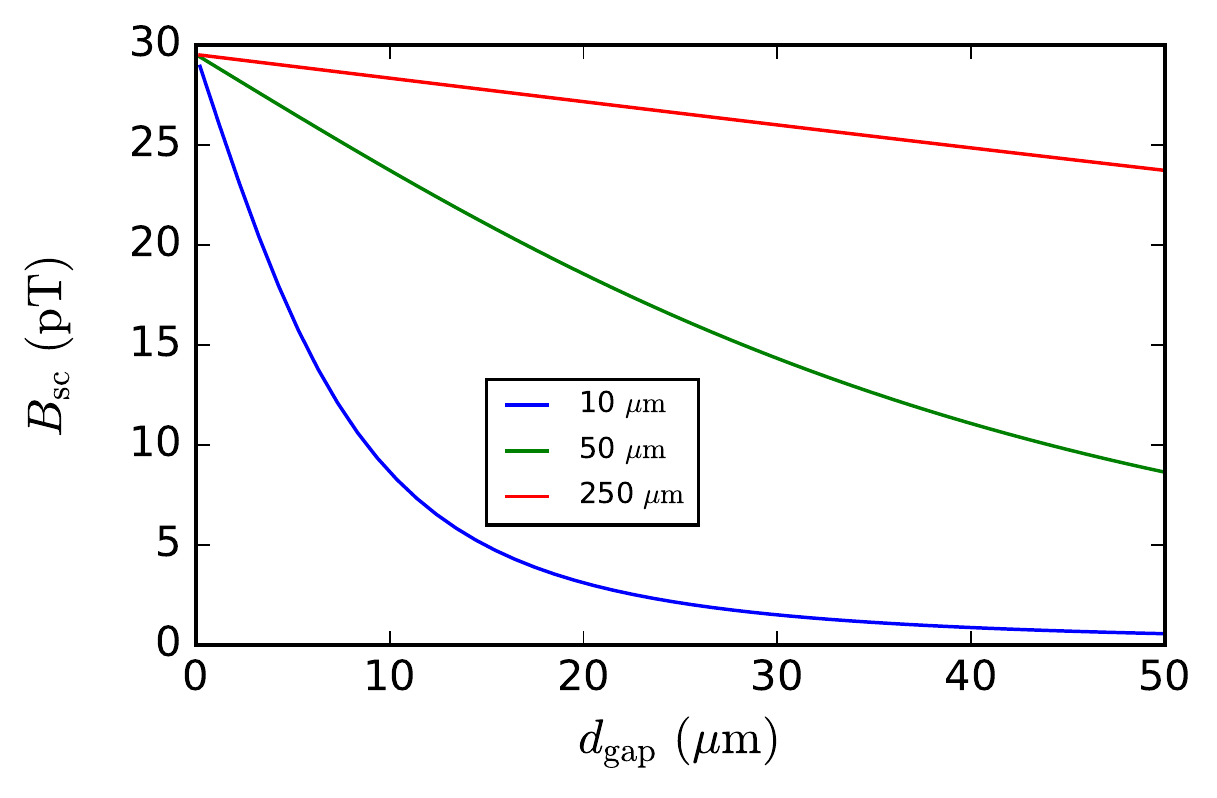}
    \caption{\textbf{Magnetic field from moving surface charges}. The magnetic field component along the NV sensing axis due to surface charges on the moving test mass ($\sigma=10~{\rm \upmu C/m^2}$, $v=4.7~{\rm m/s}$) is plotted as a function of $d_{\rm gap}$ for three different test mass radii (denoted in the legend).}
    \label{fig:surface charge current}
\end{figure}

This source of systematic error is important and merits further study. We cannot rule out that it contributed to the non-zero result claimed in Ref.~\cite{rong:2020observation}. However, the interaction effective field amplitude reported in that work, $\gtrsim10~{\rm nT}$, is expected to be an order of magnitude larger than this effect.

Fortunately, as shown in Fig.~\ref{fig:geometryopt}(d), the scaling of the exotic interaction signal is $\propto e^{-d_{\rm gap}}$. On the other hand, the spurious magnetic field due to moving surface charge falls off much slower with $d_{\rm gap}$. To analyze the dependence, we assume the test mass charge is uniformly distributed over a finite circle (radius: $R_{\rm tm}$) on the face of the test mass parallel to the diamond (distance: $d_{\rm gap}$). The magnitude of the magnetic field transverse to $\bm{v}$ from the associated surface current is given by:
\begin{align}
    B_{\rm sc} = \frac{\mu_0 J}{4\pi}\int_{0}^{R_{\rm tm}}\int_0^{2 \pi} \frac{\rho~ d_{\rm gap}}{(\rho^2+d_{\rm gap}^2)^{3/2}}d\phi d\rho,
\end{align}
where $J=\sigma v$ is the peak surface current density. The result from integration gives:
\begin{align}
    B_{\rm sc}= \frac{\mu_0 J}{2} \left(1 - \frac{d_{\rm gap}}{\sqrt{R_{\rm tm}^2 + d_{\rm gap}^2}} \right).
    \label{eq:surfaceB}
\end{align}

Figure~\ref{fig:surface charge current} shows $B_{\rm sc}$ as a function of $d_{\rm gap}$ for different test mass dimensions, calculated using Eq.~\eqref{eq:surfaceB}. For small test mass radii, $R_{\rm tm}\ll d_{\rm gap}$, we find $B_{\rm sc}\propto (R_{\rm tm}/d_{\rm gap})^2$. For larger test mass radii $R_{\rm tm}\gg d_{\rm gap}$, $B_{\rm sc}$ is nearly constant. In either case, the field falls off more slowly than $e^{-d_{\rm gap}}$. This may enable independent studies of this systematic effect and allow for methods of cancellation.

\subsection{Dielectric moving in magnetic field}
A dielectric moving in a static magnetic field becomes polarized. This polarization induces surface charges on the material which in turn produces a spurious magnetic field transverse to the axis of motion. As for the previous case of moving surface charges, this will effect experiments probing potentials $V_{4+5}$,$V_{14}$, and $V_{15}$. 

At sufficiently low velocities ($v \ll c$) the electric field in a boosted frame moving relative to another is given by:  
\begin{align}
    \label{eq:b-field_boost}
    \vec{E}' = \vec{E} + \vec{v} \times \vec{B}. 
\end{align}
This field, $\vec{E'}$, will polarize the dielectric test mass, inducing a surface charge. If we assume a spherical test mass and that $\vec{E}=0$, then the induced surface charge is given by:
\begin{align}
    \label{eq:polarization}
    \sigma &= (\epsilon-\epsilon_0)\vec{n}\cdot \vec{E_{\rm{in}}} \\ 
           &= \frac{(\epsilon-\epsilon_0)\epsilon_0}{\epsilon_0 + N_{\rm tm}(\epsilon-\epsilon_0)}\vec{v}\times\vec{B},
\end{align}
where $\vec{E_{\rm in}}$ is the electric field inside the test mass, $N$ is the depolarization factor and $\epsilon$ is the permittivity of the test mass material. The largest permittivity material considered here is diamond, $\epsilon\approx5.5\epsilon_0=5\times10^{-11}~{\rm F/m}$. Setting $v = 4.7~{\rm m/s}$ and $B_0 = 10~{\rm mT}$, and assuming (for simplicity) a spherical test mass ($N_{\rm tm}=1/3$), we find the surface charge from this effect is $\sigma\approx 10^{-6}~{\rm \upmu C/m^2}$. The magnetic field resulting from such a surface current is well below $1~{\rm fT}$ and can be safely neglected. 

\subsection{Internal thermal motion of nuclei}
The velocity of the test mass movement considered here is ${\sim}5~{\rm m/s}$. At first glance it might seem that the internal thermal velocity of nuclei, which at room temperature is of order ${\sim}500~{\rm m/s}$ would dwarf the signal from the macroscopic movement of the test mass. However, the pulse sequence employed is sensitive to both the magnitude and the sign of the velocity, and the measurement addresses a large number of nucleons ($N\gg10^{10}$). The expectation value of the nuclear thermal motion is zero and the second order contribution scales as $N^{-1/2}t^{-1/2}$. Thus, we expect that the noise contribution due to nuclear thermal motion can be neglected.

\subsection{Magnetic field gradients}
\label{sec:gradients}
For potentials where spin polarized test masses are necessary, nuclear spin polarization is inevitably accompanied by a non-zero magnetization of the test mass. The stray magnetic fields produced by magnetized test masses couple to the NV electron spins in the same manner as effective interaction fields, and thus it is important to suppress and differentiate them.  
 
 To estimate the magnitude of this effect, we used finite-element modelling for the geometries shown in Fig.~\ref{fig:angle_adjust} for potentials $V_{6+7}$, $V_{14}$, and $V_{15}$. The magnetization of the polarized test mass can be estimated as $M=\rho\mu_{\rm C}\approx0.2~{\rm A/m}$, where $\rho=5\times10^{25}~{\rm m^{-3}}$ is the excess polarized $^{13}$C spin density and $\mu_{\rm C} = 3.5\times10^{-27}~{\rm J/T}$ is the magnetic moment of a single $^{13}$C nucleus. 
 
 \begin{figure}
    \centering
    \includegraphics[width=\columnwidth]{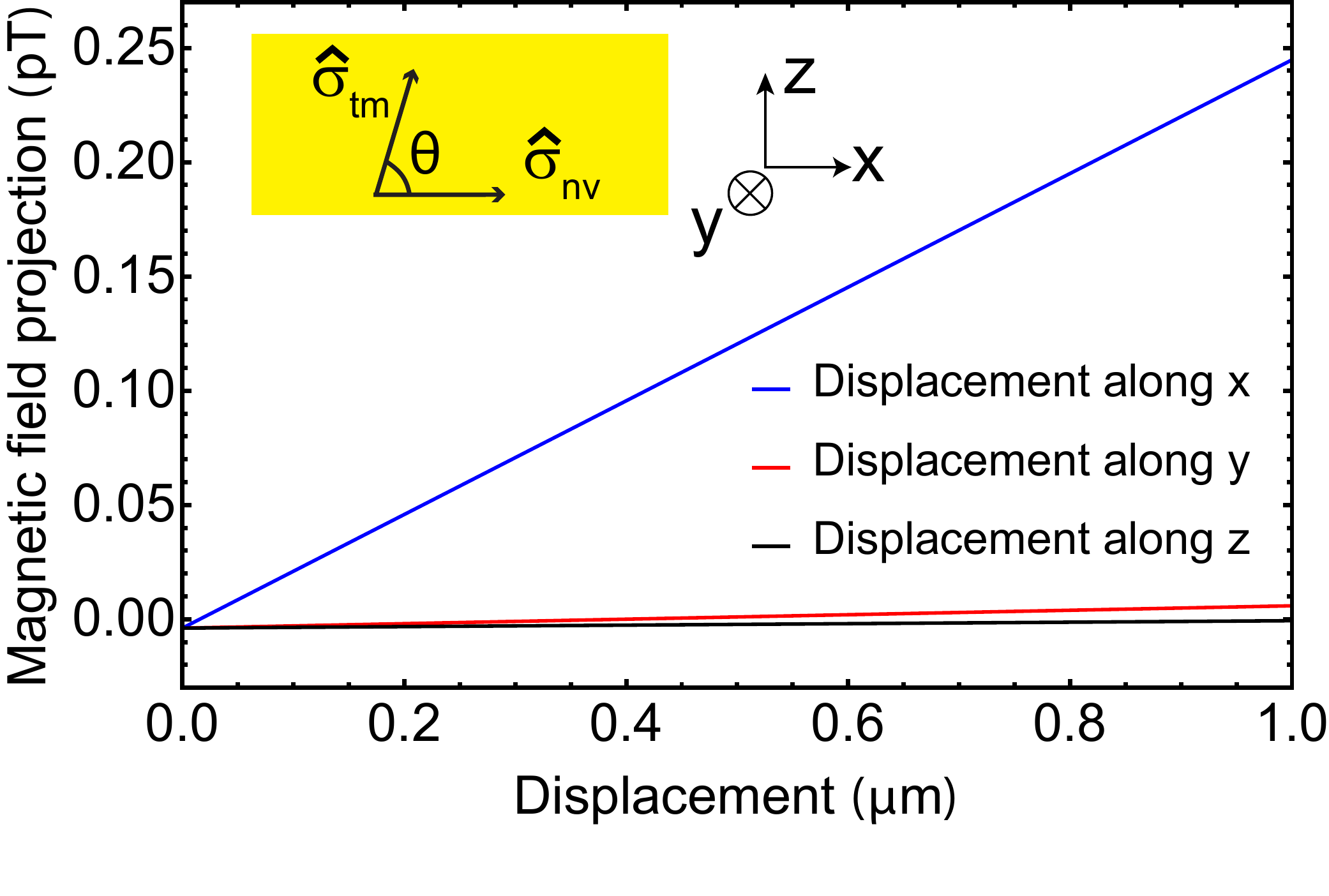}
    \caption{\textbf{Stray magnetic field from magnetized test mass.} The stray magnetic field component along the NV sensing axis, $\bm{\hat{\sigma}_{\rm nv}}$, is averaged over the NV sensing volume for different test mass displacements. The largest gradient is for test mass motion along the $\bm{\hat{\sigma}_{\rm nv}}$ axis, which is the scenario for probing $V_{6+7}$, see Fig.~\ref{fig:angle_adjust}(b).}
    \label{fig:mag_systematic}
\end{figure}

Using the experimental parameters summarized in Tab.~\ref{tab:polarized_mass}, we estimated the stray magnetic signals produced by such a magnetized test mass. Figure~\ref{fig:mag_systematic} shows the dependence of the magnetic field component along $\bm{\hat{\sigma}_{\rm  nv}}$, averaged over the NV sensing volume, as a function of displacement. For $V_{14}$ and $V_{15}$, the motion of the test mass is orthogonal to $\bm{\hat{\sigma}_{\rm nv}}$. Thus, the volume-averaged magnetic field change for small displacements ($d_1\ll R_{\rm tm}$) is zero. However, for the experimental geometry of $V_{6+7}$, the field gradient of ${\sim}250~{\rm fT/\upmu m}$ is not quite negligible. 

The main strategy to suppress this spurious magnetic signal is to exploit the fact that it is proportional to the displacement, while the exotic interaction is proportional to velocity. Since the displacement and velocity modulations are $90\degree$ out of phase, we can selectively detect the interaction signal in the appropriate quadrature signal. The detection sequence depicted in Fig.~\ref{fig:pulse}(c) is phase sensitive and can distinguish between the two signal sources, assuming the phase of the harmonic motion is known. The displacement phase may be obtained by monitoring the voltage applied to the interdigitated electrode device or by optically or capacitively detecting the test mass position. With perfect information about the test mass phase, this approach could completely cancel out the magnetic portion of the NV signal. Allowing for a small phase error $\Delta \phi$, the magnetic signal amplitude would be suppressed by a factor of $2\Delta \phi/\pi$. 

If necessary, the test mass magnetization can be directly suppressed by polarizing test-mass NV spins in the opposite direction. For a test-mass NV density of $1~{\rm ppm}$ with $90\%$ spin polarization, the maximum NV magnetization is $3.2~{\rm A/m}$. This is more than sufficient to cancel the $^{13}$C magnetization (${\sim}0.2~{\rm A/m}$). Initially NV centers are polarized in $m_s=0$, but the spins can be “tipped”  by applying resonant microwaves. The microwave field amplitude then controls the NV magnetization and can be chosen to cancel the magnetization of the $^{13}$C nuclei.

\section{Conclusion}
In this paper, we considered the potential for using diamond NV centers to probe new spin interactions at the micrometer scale. We analyzed five hypothetical potentials characterizing two possible spin-velocity interactions ($V_{12+13}$ and $V_{4+5}$) and three velocity-dependent spin-spin interactions ($V_{6+7}$, $V_{14}$, $V_{15}$). We considered several sources of systematic errors, described how they may affect the measurement of the exotic interactions, and outlined potential mitigation strategies. For most of the interactions, our proposed experiments could improve on existing experimental constraints by 5 orders of magnitude.

\section{Acknowledgements}
The authors acknowledge valuable conversations with B. Richards, D. Budker, D. Jackson-Kimball, and A. Sushkov. This work was funded by NSF grants DMR-1809800 and CHE-1945148 and by the US DOE through the LANL/LDRD program.

\end{document}